\documentclass[aps,twocolumn,preprintnumbers,floatfix,showpacs,prx]{revtex4-1}

\usepackage{amsmath,amssymb}
\usepackage{dcolumn}
\usepackage{braket}
\usepackage{dsfont}
\usepackage{feynmp}
\usepackage{hhline}
\usepackage{graphicx}
\usepackage{mathtools}
\usepackage{xfrac}
\usepackage{paralist}
\usepackage{enumitem}
\usepackage[normalem]{ulem}
\usepackage{color}
\definecolor{darkblue}{rgb}{0.,0.,0.4}
\definecolor{darkred}{rgb}{0.5,0.,0.}

\newcolumntype{L}{>{\centering\arraybackslash}m{12.4cm}}

\newcommand{\FF}{\mathbb{F}}
\DeclareMathOperator{\coker}{\mathrm{coker}}
\DeclareMathOperator{\codim}{\mathrm{codim}}
\DeclareMathOperator{\im}{\mathrm{im}}
\DeclareMathOperator{\rank}{\mathrm{rank}}
\begin{document}

\title{Fracton Topological Order, Generalized Lattice Gauge Theory and Duality}
\author{Sagar Vijay}
\author{Jeongwan Haah}
\author{Liang Fu}
\affiliation{Department of Physics, Massachusetts Institute of Technology,
Cambridge, MA 02139, USA}
\begin{abstract}
We introduce a generalization of conventional lattice gauge theory to describe fracton topological phases,
which are characterized by immobile, point-like topological excitations, and sub-extensive topological 
degeneracy. We demonstrate a duality between fracton topological order and interacting spin systems with
symmetries along extensive, lower-dimensional subsystems, which may be used to systematically search 
for and characterize fracton topological phases. 
Commutative algebra and elementary algebraic geometry provide an effective
mathematical toolset for our results.  
Our work paves the way for identifying possible material realizations of fracton topological phases.
\end{abstract}
\maketitle

Topological phases of matter are currently attracting tremendous interest from diverse disciplines such as
theoretical physics, quantum information and quantum materials. 
Gauge theory provides a unified framework for understanding several important topological phases, from
quantum Hall states to quantum spin liquids \cite{Wegner, FradkinShenker, ZhangHassonKivelson, WenNiu,
Wen, SachdevRead, SenthilFisher, MoessnerSondhi,Kitaev, Hermele, Wenbook}.
The fractional statistics of topological excitations \cite{Halperin, Wilczek}, or anyons, can be understood from
the Aharanov-Bohm phase of a charge moving around a flux.  The topological degeneracy of the ground
state is characterized by the holonomy of a locally flat gauge connection around non-contractible loops.  

Recently, a new kind of topological phase which does not fit into the framework of 
gauge theory has been discovered in exactly solvable lattice models in three dimensions \cite{Haah,
Chamon, Bravyi, Yoshida, Fracton}.  A remarkable property of this new phase is the existence of point-like
fractional excitations termed ``fractons'' \cite{Fracton}, which can only be created at the corners of
membrane- or fractal-like operators, unlike anyons that are created at the ends of Wilson lines.  The creation of anyons at the two ends of a Wilson line immediately implies that anyons can move by repeated application of a local, line-like operator. In contrast, the
absence of any operator that can create a pair of fractons implies that a single fracton cannot
move without creating additional excitations, i.e. fractons are fundamentally \emph{immobile}. 
Thus far, two broad classes of fracton topological orders have been
found:
\begin{description}[before={\renewcommand\makelabel[1]{\bfseries ##1}}]
\item[\hspace{.2in}{Type I}] fracton phases, such as the Chamon-Bravyi-Leemhuis-Terhal (CBLT) model
\cite{Chamon, Bravyi} and the Majorana cubic model \cite{Fracton}, have fracton excitations appearing at
the corners of \emph{membrane} operators; composites of fractons form topological excitations that are only
mobile within lower-dimensional subsystems.

\item[\hspace{.2in}{Type II}] fracton phases, such as Haah's code \cite{Haah} and related models
\cite{Yoshida}, have fracton excitations that appear at the corners of \emph{fractal} operators.  {All}
topological excitations are strictly localized and there are no mobile topological quasiparticles.
\end{description}

Fracton topological order provides an exciting development in the search for new quantum phases of matter,
for new schemes for quantum information processing \cite{Terhal}, and in the investigation of glassy
dynamics in interacting quantum systems \cite{Kim_Haah}. Fractons enable new forms of electron
fractionalization \cite{Fracton}, and provide an alternative to Fermi or Bose statistics in three dimensions. 
Fractons may be used to build a robust, finite-temperature quantum memory, as theoretically demonstrated
for Haah's code \cite{Haah, Bravyi_Haah}.  The innately slow dynamics of fractons provides an intriguing
connection with quantum glasses, many-body localization, and a new testing ground for the postulates of
quantum statistical mechanics.

Research on fracton topological phases is in its early stages and has been based on studies of specific
lattice models.
It is thus highly desirable to find a more unified theoretical framework for fracton topological order.  
In this work, we demonstrate that fracton topological phases can be obtained as the quantum dual of $d$-dimensional systems that possess ``subsystem symmetries", namely a set of symmetries associated with
subsystems of dimension $2 \le d_{s} < d$.
Specifically, we establish an exact duality relating both type-I and type-II fracton topological orders 
to symmetry-breaking order in quantum systems with subsystem symmetries along planes and fractals,
respectively. This duality between fracton topological order and subsystem symmetry-breaking, hereafter
referred to as the ``\emph{F-S}'' duality,  is naturally obtained from a generalized lattice gauge theory which
we introduce.   
Instead of placing a gauge field on links between neighboring sites as in a standard lattice gauge theory, 
we introduce a new field to mediate {\it multi-body} interactions between matter fields on a cluster of
neighboring sites. 
This yields an interacting quantum system with a generalized Gauss' law that characterizes the fracton
topological phase. 

Before describing our construction in generality, we present a concrete example that yields a new class of 
type-I fracton topological phases.  Consider a model of Ising spins at the sites of a three-dimensional cubic 
lattice, whose Hamiltonian ($H_{\mathrm{plaq}}$) is defined to be a sum of four-spin interactions at each 
plaquette, as shown in Table \ref{fig:Classical_Spins}.  This classical ``plaquette Ising model" is invariant 
under a spin-flip $\tau\rightarrow -\tau$ along any $xy$, $yz$ or $xz$-plane of the cubic lattice.  The 
plaquette Ising model has a rich history of study, attracting interest as a model for the statistical mechanics 
of smooth surfaces, and as a lattice regularization of string theory \cite{Savvidy_Wegner, Savvidy_2, 
Savvidy_3, Wegner_2, Ising_Plaquette_Slow_1, Ising_Plaquette_Slow_2}.

We introduce a generalized lattice gauge theory to 
construct the quantum dual of the plaquette Ising model in a transverse field. This generalizes Wegner's duality 
\cite{Wegner} between the $d$-dimensional transverse-field Ising model and Ising lattice gauge theory 
\cite{FradkinShenker}.  Wegner's duality is motivated by the observation that a configuration of Ising spins 
may be specified by the locations of the \emph{domain walls} between symmetry-breaking states of the 
Ising model. As a result, a dual representation of the Ising matter is given by Ising ``domain wall" fields on 
the links of the lattice.  
Furthermore, since domain walls form closed, ($d-1$)-dimensional surfaces, physical states in the domain 
wall Hilbert space must satisfy a local ``zero-flux'' condition, that the lattice curl of the domain wall spins 
vanishes around each plaquette.   In this way, the $d$-dimensional transverse-field Ising model is dual to $
\mathbb{Z}_{2}$ lattice gauge theory.  

\begin{figure}
$\begin{array}{cc}
  \includegraphics[trim = 0 0 0 0, clip = true, width=0.25\textwidth, angle = 0.]{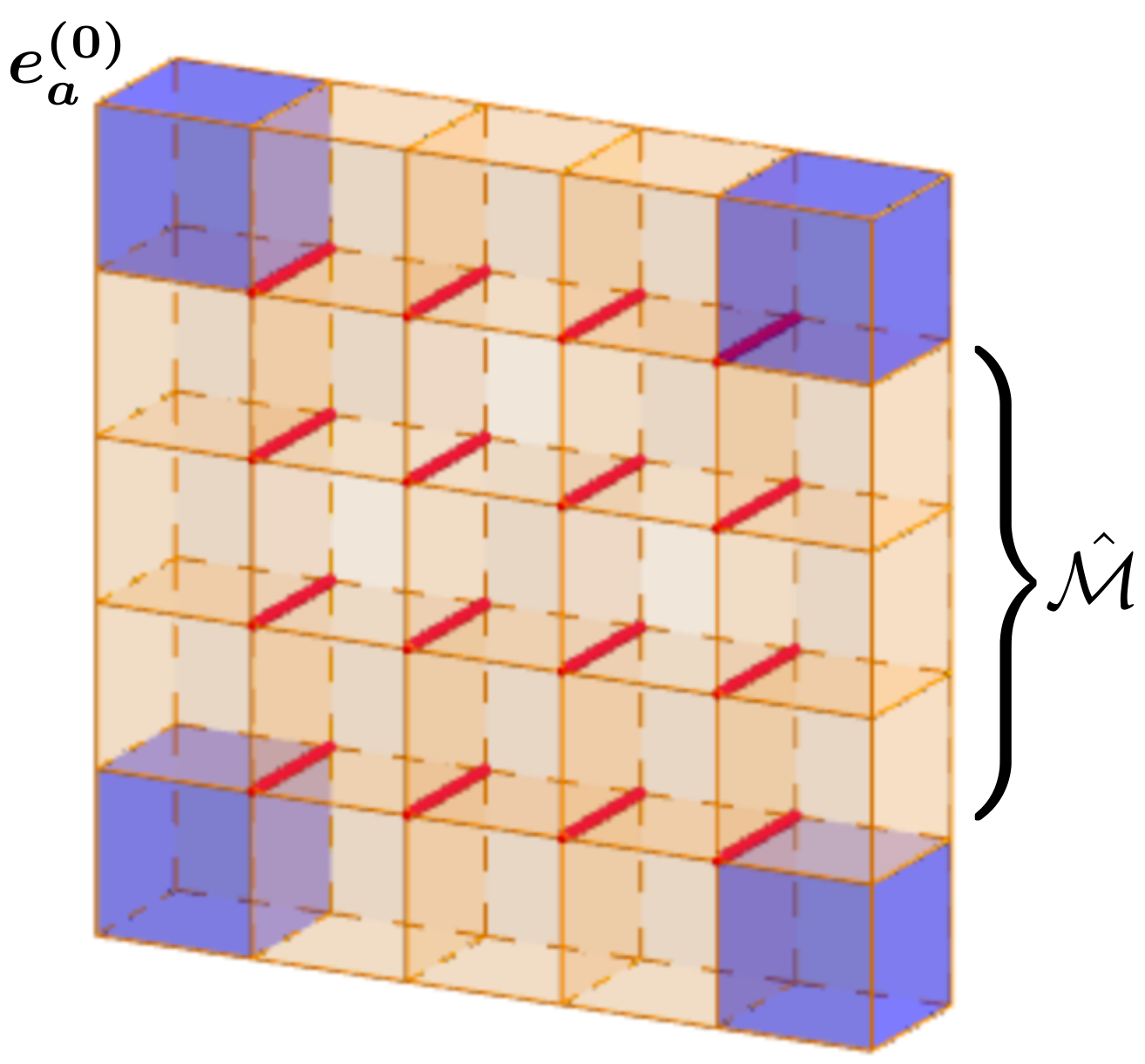} &
   \includegraphics[trim = 0 0 0 0, clip = true, width=0.23\textwidth, angle = 0.]{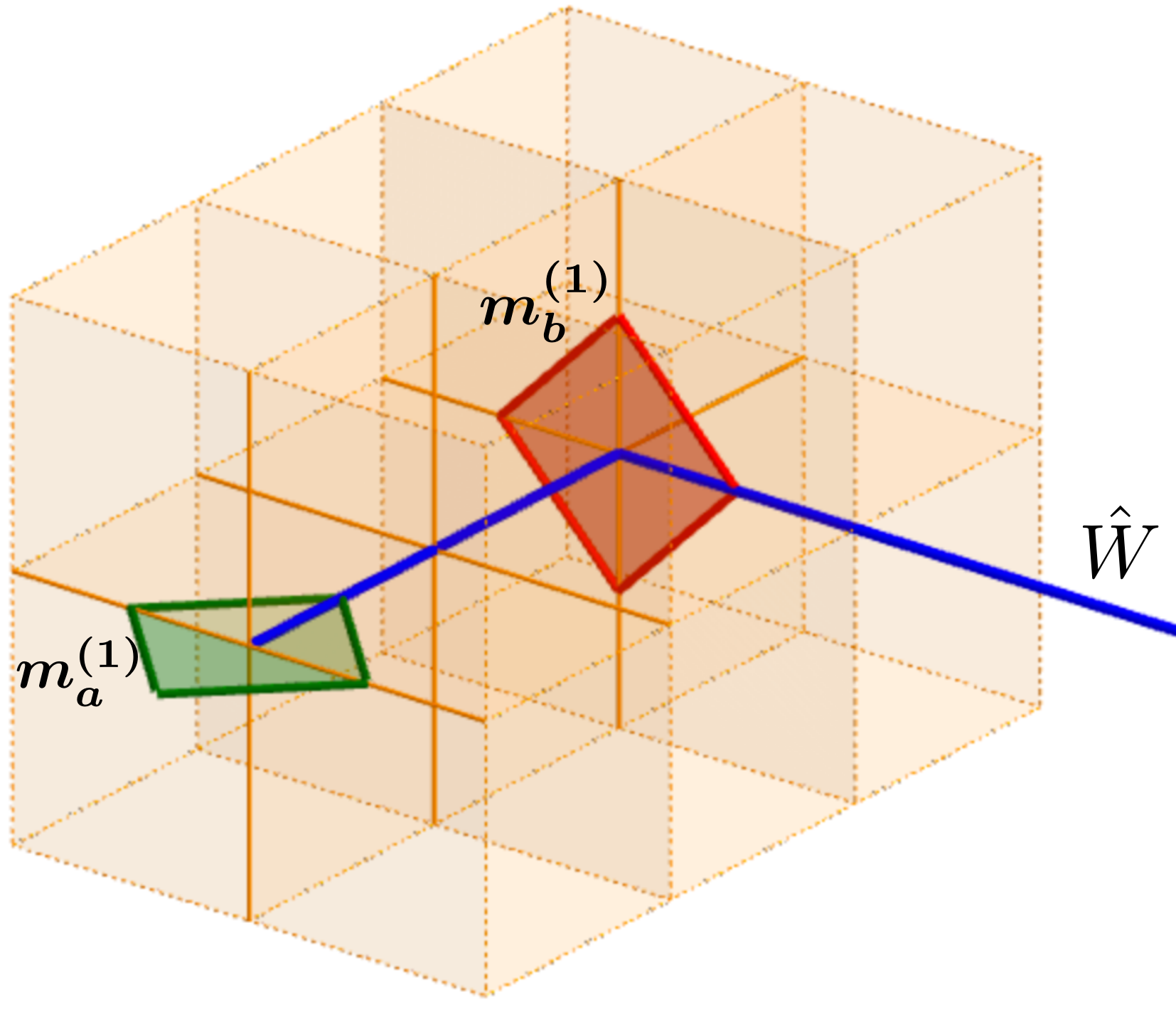}\\
 \text{(a)} & \text{(b)}
  \end{array}$
 \caption{The fundamental excitations of the X-cube model are shown in (a) and (b). Acting on the ground-
 state of the X-cube model with a product of $\sigma^{z}$ operators along the colored red links that lie within a flat, rectangular region $\mathcal{M}$ 
 generates four fracton cube excitations ($e^{(0)}_{a}$) at the corners of the region.  A straight Wilson line of $\sigma^{x}$ 
 operators acting on the blue links in (b) isolates a pair of quasiparticles ($m^{(1)}_{a}$ or $m^{(1)}_{b}$) at the ends, that are only free to move along the line.   Attempting to move these quasiparticles in any other direction by introducing a corner in the Wilson line, creates a topological excitation at the corner as shown in (b).}  
  \label{fig:X_Cube}
\end{figure}

Our duality between fracton topological order and subsystem symmetry-breaking is obtained by a similar 
observation. A configuration of Ising spins may, equivalently, be specified by the eigenvalue of each 
\emph{interaction term} in the Hamiltonian \cite{Footnote}. For example, to obtain the dual of the plaquette 
Ising model, we are naturally led to introduce the Ising fields $\{\sigma\}$ at the center of each plaquette.  
Physically, the $\sigma$ field labels the presence or absence of a domain wall between the subsystem 
symmetry-breaking ground-states of the plaquette Ising Hamiltonian.  While domain walls in the ordinary 
Ising model form closed surfaces, the $\sigma$ fields in our model must satisfy more exotic local 
constraints due to the geometry of the plaquette interactions to ensure a one-to-one correspondence with 
the physical space of domain walls in $H_{\mathrm{plaq}}$.

As we demonstrate below, the quantum dual of the plaquette Ising Hamiltonian, in terms of the $\sigma$ 
fields, exhibits fracton topological order.  The resulting fracton Hamiltonian has a solvable limit, analogous to 
the deconfined phase of a conventional gauge theory, which is  given by a Hamiltonian for the Ising fields ($
\sigma$), now placed at the links of the dual cubic lattice.  As shown in Table \ref{fig:Classical_Spins}, this 
fracton Hamiltonian consists of two types of terms: (1) a twelve-spin $\sigma^{x}$ interaction for the spins 
surrounding a dual cube and (2) four-spin $\sigma^{z}$-interactions at each vertex of the dual cubic lattice 
that are aligned along the $xy$, $yz$ and $xz$-planes.  The cubic and cross-like geometries of the 
interactions motivate the name ``X-cube" model.  The ground-state is topologically-ordered, as the ground-states are locally indistinguishable, and one of the fundamental excitations -- obtained by flipping the 
eigenvalue of the cubic interaction term -- is a fracton.  This can be seen by observing that there is no local 
operator that can create a single pair of cube excitations.  For example, the operator $\sigma^{z}_{n}$ 
creates four cube excitations when acting on the ground-state.  Repeated application of $\sigma^{z}_{n}$ over a membrane separates the four cube excitations to each \emph{corner} as shown in Figure 
\ref{fig:X_Cube}a.  Therefore, a single cube excitation is 
fundamentally immobile, and cannot move without creating additional cube excitations. Pairs of cube excitations, however, can be moved by sequentially applying a local, membrane-like operator.

The quasiparticle content of the X-cube model is summarized in Table \ref{fig:Classical_Spins}, along with 
other fracton phases such as Haah's code, the CBLT model, and a new spin model which we introduce and term the ``checkerboard model".  All of these phases are obtained by applying our generalized lattice gauge theory prescription to spin models with subsystem 
symmetries.  As we will demonstrate, a simple property of the classical spin model, that no product of 
interaction terms acts   exclusively on a pair of isolated spins, guarantees that its quantum dual exhibits 
fracton topological order.  

\begin{table*}
\begin{tabular}{|c|c|c|}
  \hline
  $\begin{array}{c}
  \\
  \textbf{\underline{Classical Spin System}}\\
  \,
  \end{array}$ & $\begin{array}{c}
  \\
  \textbf{\underline{Subsystem Symmetry}}\\
  \,
  \end{array}$ & $\begin{array}{c}
  \\
  \textbf{\underline{Fracton Topological Phase}}\\
  \,
  \end{array}$\\
  \hline
 $\begin{array}{c}
   \parbox[c][1.1in]{2.2in}{\includegraphics[trim = 207 228 171 190, clip = true, width=0.3\textwidth, angle = 0.]{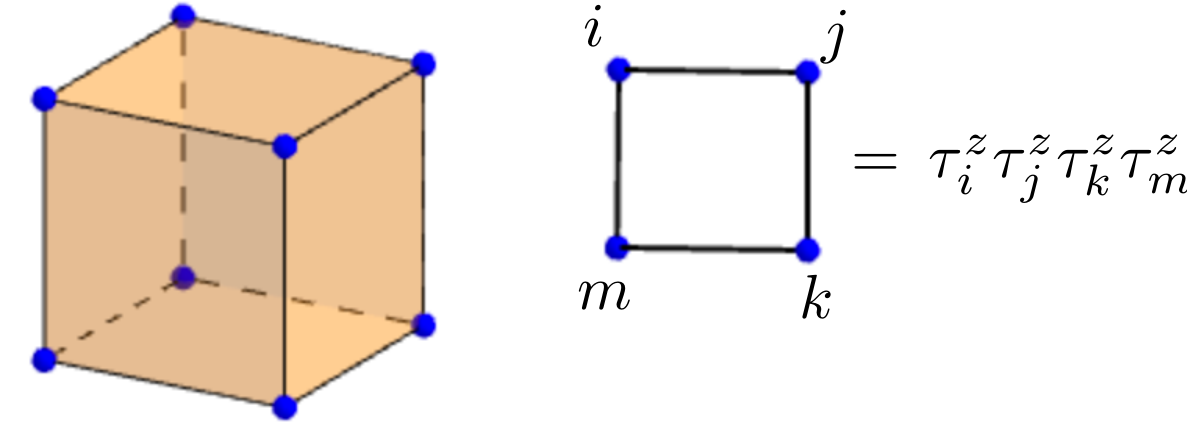}}\\\\
   \text{\bf Plaquette Ising Model}
   \end{array}$
   & 
   $\begin{array}{c}
   \parbox[c][1.2in]{1.in}{\includegraphics[trim = 0 0 0 0, clip = true, width=0.17\textwidth, angle = 0.]{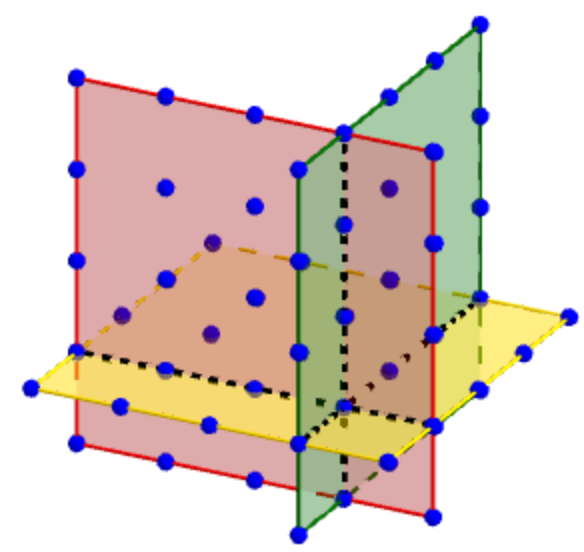}}\\
   \text{ Planar}\\
   \,
   \end{array}$
   & 
   $\begin{array}{c}
\\
   \parbox[c][1.1in]{2.1in}{\includegraphics[trim = 0 0 0 0, clip = true, width=0.29\textwidth, angle = 0.]{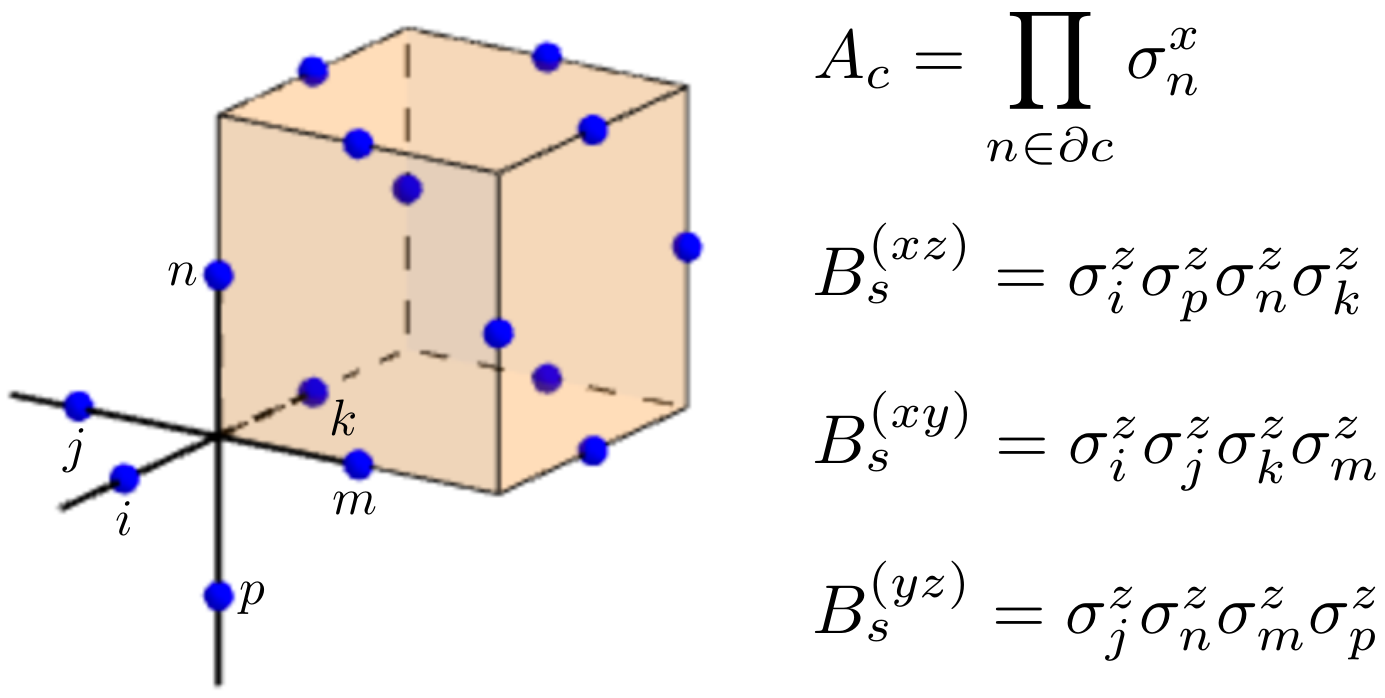}}\\\\
   \begin{array}{c}
   \text{{\bf X-Cube Model}}\\\\
   $\Big[$\text{{Type I}}:\,\,e_{a}^{(0)},\,
   m^{(1)}_{a},\, m^{(1)}_{b}\Big]
   \end{array}\\
   \,
   \end{array}$\\
   \hline
   $\begin{array}{c}
   \parbox[c][1.1in]{2.2in}{\includegraphics[trim = 205 372 192 7, clip = true, width=0.28\textwidth, angle = 0.]{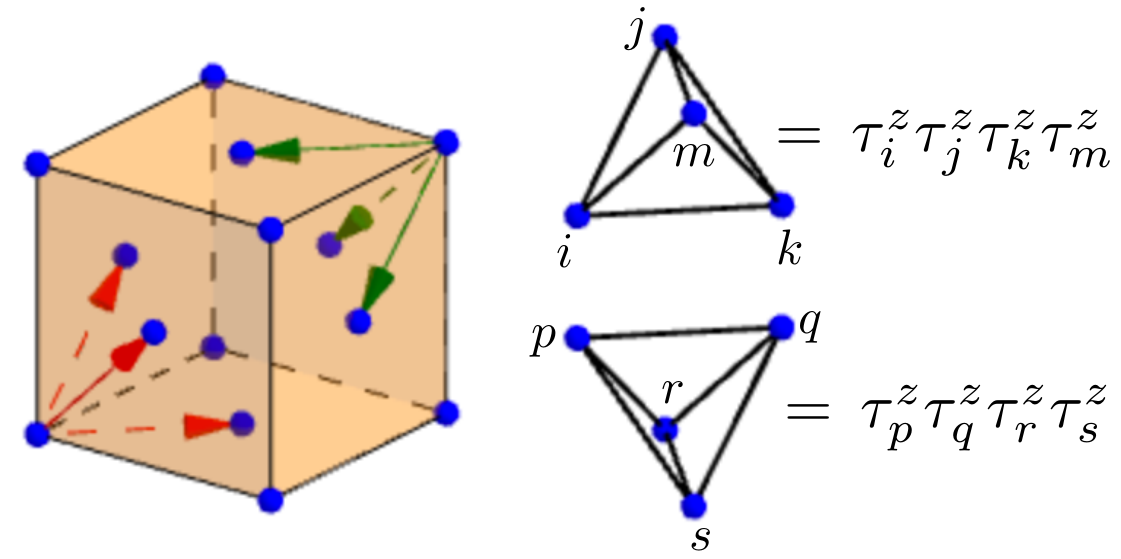}}\\
   \\
   \text{\bf Tetrahedral Ising Model}\\
   \,
   \end{array}$
   &
   $\begin{array}{c}
   \parbox[c][1.2in]{1.in}{\includegraphics[trim = 0 0 0 0, clip = true, width=0.17\textwidth, angle = 0.]{Planar_Symm_2}}\\
   \\
   \text{ Planar}\\
   \,
   \end{array}$
   & 
   $\begin{array}{c}
   \\
   \parbox[c][1.1in]{2.5in}{\includegraphics[trim = 0 0 0 0, clip = true, width=0.25\textwidth, angle = 0.]{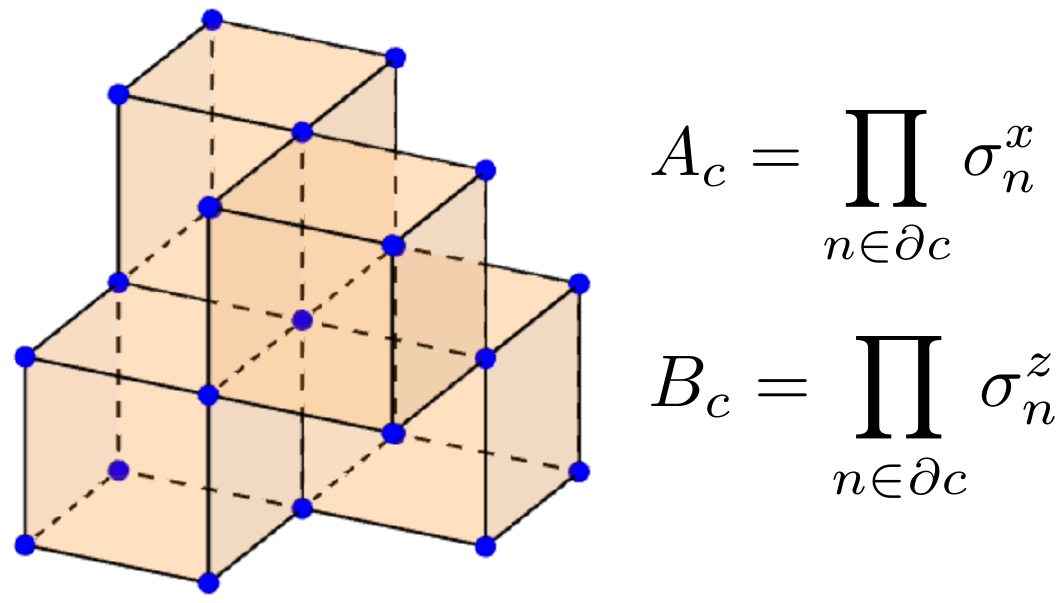}}\\
   \\
   \begin{array}{c}
   \text{\bf Checkerboard Model}\\\\
   $\Big[$\text{{Type I}}:\,\,e_{a}^{(0)},\, m^{(0)}_{a}\Big]
   \end{array}\\
   \,
   \end{array}$\\
   \hline
   $\begin{array}{c}
   \parbox[c][1.3in]{2.2in}{\includegraphics[trim = 0 0 0 0, clip = true, width=0.28\textwidth, angle = 0.]{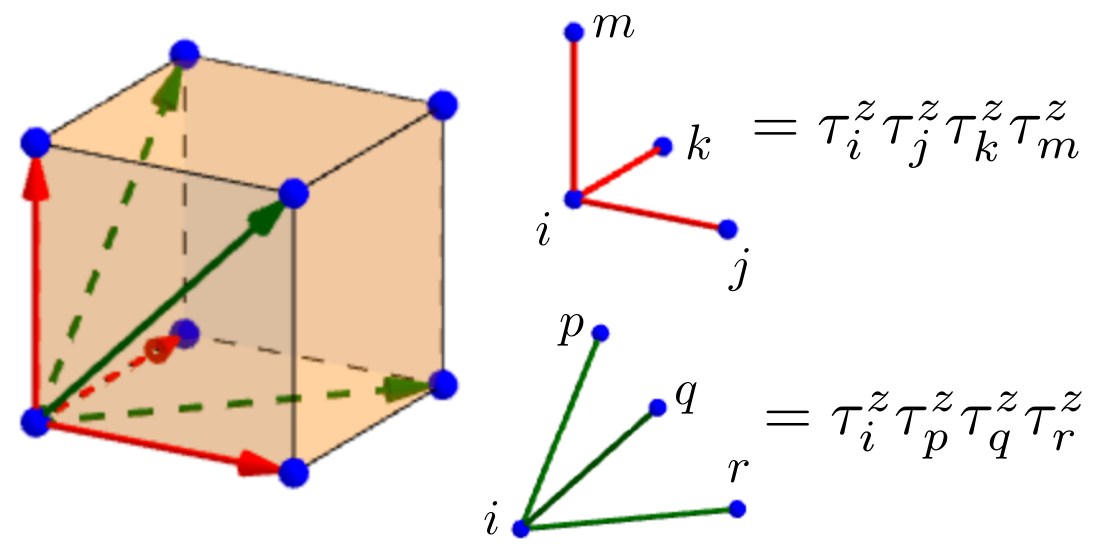}}\\
   \\
   \text{\bf Fractal Ising Model}\\
   \,
   \end{array}$
   &
   $\begin{array}{c}
   \parbox[c][1.3in]{1.2in}{\includegraphics[trim = 10 0 0 10, clip = true, width=0.17\textwidth, angle = 0.]{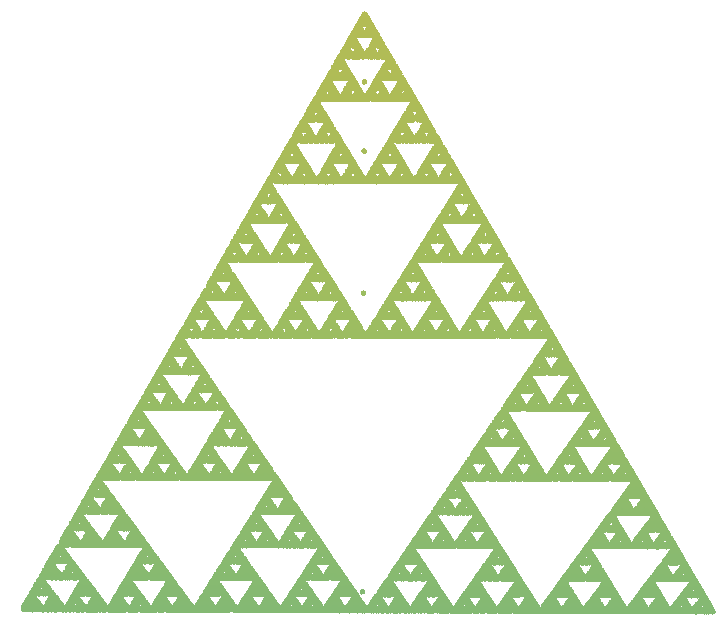}}\\
   \\
   \text{Fractal}\\
   \,
   \end{array}$
   & 
   $\begin{array}{c}\\
   \parbox[c][1.42in]{2.75in}{\includegraphics[trim = 0 0 0 0, clip = true, width=0.34\textwidth, angle = 0.]{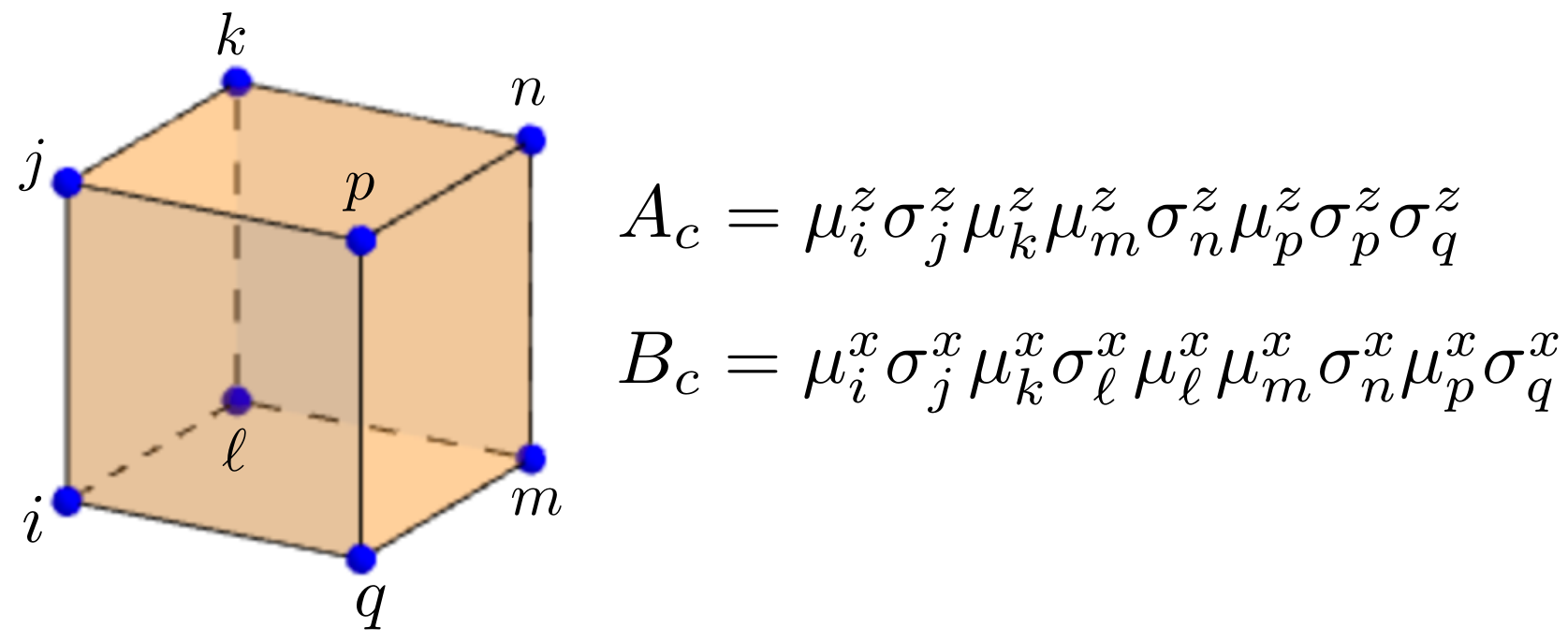}}\\
   \begin{array}{c}
   \text{\bf Haah's Code}\\\\
   $\Big[$\text{{Type II}}:\,e_{a}^{(0)},\,m_{a}^{(0)}\Big]
   \end{array}\\
   \,
   \end{array}$\\
   \hline
   \end{tabular}
 \caption{Representative examples of fracton topological orders built from classical spin systems with 
 subsystem symmetry.  The classical models shown above are the plaquette, tetrahedral, and fractal Ising 
 models ($H_{\mathrm{plaq}}$, $H_{\mathrm{tetr}}$, and $H_{\mathrm{frac}}$, respectively).  In the 
 plaquette Ising model, spins are placed on the sites of a simple cubic lattice as shown, and the Hamiltonian 
 is a sum of four-spin interactions at the face of each cube. In the tetrahedral Ising model, spins are 
 arranged on an fcc lattice, and each spin participates in four-spin interactions coupling neighboring spins 
 that form a tetrahedron, as indicated.  Finally, the fractal Ising model consists of two types of four-spin 
 interactions at each cube.  The model has a fractal symmetry, and is invariant under a spin-flip along a 
 three-dimensional Sierpinski triangle, as elaborated in Appendix A.  The X-cube, checkerboard, and Haah's 
 code fracton models ($H_{\mathrm{X}\text{-}\mathrm{Cube}}$, $H_{\mathrm{check}}$, and 
 $H_{\mathrm{Haah}}$, respectively) are solvable limits of fracton phases that are obtained by promoting the 
 subsystem symmetry of the indicated spin models to a local symmetry.  The fracton model $H_{\mathrm{X}
 \text{-}\mathrm{Cube}}$ is naturally represented by placing spins $\sigma$ on links of the cubic lattice, and 
 is a sum of a twelve-spin $\sigma^{x}$-operator at each cube and the indicated planar, four-spin $
 \sigma^{z}$-operators. The checkerboard model is a sum of eight-spin $\sigma^{x}$- and $\sigma^{z}$-
 interactions over cubes arranged on an fcc lattice and is self-dual under the exchange $\sigma^{x}
 \longleftrightarrow\sigma^{z}$. Only the fundamental excitation types are indicated above. Here, we have 
 adopted the notation $e_{a}^{(j)}$ ($m_{a}^{(j)}$) to refer to a dimension-$j$ excitation -- i.e. an excitation 
 that is only free to move within a dimension-$j$ subsystem without creating additional excitations -- that is 
 obtained by flipping  a $\sigma^{x}$-type ($\sigma^{z}$-type) interaction. For type-I fracton phases, 
 bound-states of fracton 
 excitations can form mobile quasiparticles; these mobile composite excitations are not indicated in the 
 table.}
 \label{fig:Classical_Spins}
\end{table*}

More generally, consider a {classical} Hamiltonian for Ising spins $(\tau_{i})$ at the sites of a three-dimensional Bravais lattice.  We assume, for simplicity of presentation, that there is a single spin at each 
lattice site; the case where the unit cell is larger is explained in the Appendix. The Hamiltonian consists 
of $\ell$ types of interactions at each lattice site $i$, and may be written in the form:
\begin{align}\label{eq:H_0}
H_{0} = -t\sum_{i}\left(\mathcal{O}^{(1)}_{i}[\tau] + \cdots + \mathcal{O}^{(\ell)}_{i}[\tau]\right).
\end{align}
with the constant $t > 0$. We demonstrate that a classical spin Hamiltonian (\ref{eq:H_0}) satisfying certain 
simple properties may be used to build a topologically-ordered, quantum system with {fracton} excitations.  
First, we require that the spin Hamiltonian (\ref{eq:H_0}) has a \emph{subsystem symmetry} under which the 
spin-flip transformation $\tau \rightarrow -\tau$ along non-local {subsystems} of the lattice -- i.e. subsystems 
that scale with the system size  -- leaves $H_{0}$ invariant.  We further require that $H_{0}$ has no 
\emph{local} symmetries.  In this sense, a subsystem symmetry is ``intermediate" between local and global symmetries \cite{Nussinov_Batista, Nussinov_Fradkin}.  For the remainder of this work, we will refer to the plaquette Ising model 
($H_{\mathrm{plaq}}$) and the tetrahedral Ising Hamiltonian $(H_{\mathrm{tetr}})$ as concrete examples.  
As shown in Table \ref{fig:Classical_Spins}, the Hamiltonian $H_{\mathrm{tetr}}$ is defined on the face-
centered cubic (fcc) lattice and consists of nearest-neighbor four-spin interactions that form elementary 
tetrahedra. The tetrahedral Ising model has two interaction terms per site on the fcc lattice. Both the tetrahedral and plaquette Ising models have a subsystem symmetry, as they are 
invariant under spin-flips along orthogonal planes ($xy$, $yz$ or $xz$). 

An important consequence of the subsystem symmetry of the spin Hamiltonian $H_{0}$ is that the resulting 
ground-state has sub-extensive \emph{classical} degeneracy $D$, taking the form $\log\,D \sim O(L)$  on 
the length-$L$ three-torus.  Since the degeneracy is classical in nature, each ground-state may be 
distinguished by a local order parameter.  Transitioning between ground-states, however, requires 
performing a spin-flip along a subsystem.  While a local perturbation can lift the classical degeneracy, no 
local operator can connect distinct ground-states.

\section{Generalized Lattice Gauge Theory and the \emph{F-S} Duality} 

We now build a quantum Hamiltonian 
with fracton topological order  by promoting the subsystem symmetry of the spin system (\ref{eq:H_0}) to a 
local symmetry.  We begin by adding a transverse field at each lattice site 
to allow the classical spins to exhibit quantum fluctuations. 
Next, we introduce additional Ising spins $(\sigma_{i,a})$ at the center of each multi-spin interaction 
appearing in $H_{0}$; these spins appear at the sites of a lattice with an $\ell$-site basis. 
We introduce a minimal coupling $H_{c}$ between the $\sigma$ spins and the Ising matter fields, by 
coupling each $\sigma$ to its corresponding multi-spin interaction:
\begin{align}
H_{\mathrm{c}} \equiv - t \sum_{i}\left(\sigma^{z}_{i,1}\,\mathcal{O}^{(1)}_{i}[\tau^{z}] + \cdots + \sigma^{z}_{i,\ell}\,\mathcal{O}^{(\ell)}_{i}[\tau^{z}]\right).
\end{align}
After this minimal coupling, the Hamiltonian describing both $\sigma$ and Ising matter fields is given by 
\begin{align}\label{eq:H_1}
H = H_{c} - h\sum_{i}\tau^{x}_{i}. 
\end{align}
We refer to $\sigma$ as the \emph{nexus field}, as each $\sigma$ is placed at the center of an elementary 
multi-spin interaction of the classical spin Hamiltonian.  We will soon observe that the nexus field provides a 
natural generalization of a gauge field in a conventional lattice gauge theory.  
In contrast to our construction, applying the standard gauging procedure to any of the spin models shown in 
Table \ref{fig:Classical_Spins} by introducing a gauge field on the links of the cubic lattice would result in a 
Hamiltonian with conventional $\mathbb{Z}_{2}$ topological order.  Our procedure is also distinct from 
discretizations of ``higher-form" gauge theories, in which interactions between $(n-1)$-form matter fields are 
mediated by an $n$-form gauge field \cite{Gerbe, Gerbe_2}.

The subsystem symmetry of the classical spin system (\ref{eq:H_0}) has now been promoted to a local spin-
flip symmetry of the Hamiltonian (\ref{eq:H_1}).  While $\tau^{x}_{n}$ -- the generator of a single spin-flip -- 
anti-commutes with several multi-spin interactions in (\ref{eq:H_0}), this can be compensated by acting with 
the nexus field $\sigma^{x}_{j,a}$ on the lattice sites associated with these interactions.  As a result, the 
operator
\begin{align}
&G_{i} = \tau^{x}_{i}A_{i}
\end{align}
where
\begin{align}
&A_{i} \equiv \prod_{(j,a)\in P(i)}\sigma^{x}_{j,a}
\end{align}
generates a local symmetry of the Hamiltonian ($[G_{i}, H] = 0$). The set $P(i)$ specifies the 
locations of multi-spin interactions that anti-commute with $\tau^{x}_{i}$.  

We proceed to add all other interaction terms involving the nexus field and the Ising spins that are consistent 
with this local spin-flip symmetry.  To lowest-order, we include a transverse field for the matter and nexus 
fields: 
\begin{align}\label{eq:H_ss}
H = 
- t\sum_{i,a}\sigma^{z}_{i,a}\,\mathcal{O}_{i}^{(a)}[\tau^{z}] - h\sum_{i}\tau^{x}_{i} - J\sum_{i,a}\sigma^{x}_{i,a} 
\end{align}
Since the operator $G_{i}$ generates a local symmetry of the Hamiltonian, we may impose the condition 
$G_{i}\ket{\Psi} = \ket{\Psi}$ on the Hilbert space, which amounts to a generalized Gauss's law.  In analogy 
with conventional gauge theory, we will refer to $A_{i}$ as the \emph{nexus charge} operator.  As an 
example, the nexus charge operator in the X-code model is given by the product of $\sigma^{z}$ on twelve
spins sitting at the links surrounding a cube, as shown in Table \ref{fig:Classical_Spins}.

Since the generalized Gauss's law condition commutes with the spin-nexus Hamiltonian (\ref{eq:H_ss}), it is 
possible to choose a ``gauge" that completely eliminates the Ising matter fields, and yields a Hamiltonian 
exclusively for the nexus spins.  
First, we impose the generalized Gauss's law $\tau^{x}_{i}\ket{\Psi} = A_{i}\ket{\Psi}$ to obtain a Hamiltonian 
acting within the constrained Hilbert space
\begin{align}
H = -t\sum_{i,a}\sigma^{z}_{i,a}\mathcal{O}^{(a)}_{i}[\tau^{z}] - h\sum_{i}A_{i} - J\sum_{i,a}\sigma^{x}_{i,a}
\end{align}
Since each $\tau^{z}_{i}$ operator commutes with the Hamiltonian, we may restrict our attention to states in 
the constrained Hilbert space that satisfy $\tau^{z}_{i} = +1$. This yields the gauge-fixed Hamiltonian
\begin{align}\label{eq:H_gauge_fixed}
H = - t\sum_{i,a}\sigma^{z}_{i,a} - h\sum_{i}A_{i} - J\sum_{i,a}\sigma^{x}_{i,a} 
\end{align}
When $t/h \ll 1$, it is convenient to identify an effective Hamiltonian that takes the form
\begin{align}
H_{\mathrm{eff}} = - K \sum_{i,k} B_{i}^{(k)} - h\sum_{i}A_{i} - J\sum_{i,a}\sigma^{x}_{i,a}
\end{align}
where we have introduced operators $B^{(k)}_{i}$ at each lattice site $i$. These operators are determined in 
perturbation theory by computing the simplest product of $\sigma^{z}$ terms near a given lattice site that 
commute with the nexus charge $[B^{(k)}_{i}, A_{j}] = 0$.  As an example the $B_{i}^{(k)}$ operators 
obtained by applying this construction to the plaquette Ising model are shown in Table 
\ref{fig:Classical_Spins}. 

Our proposal bears resemblance to the construction of a conventional lattice gauge theory.  First, the 
Hamiltonian for $\mathbb{Z}_{2}$ gauge theory is recovered from the general form of the Hamiltonian for the 
Ising matter and nexus fields (\ref{eq:H_ss}) if the matter fields couple through nearest-neighbor two-body 
interactions, so that $\mathcal{O}_{i}^{(a)} = \tau^{z}_{i}\tau^{z}_{i_{a}}$ where $i_{a}$ is nearest-neighbor to 
site $i$.  This is in contrast to the multi-body interactions that are present in our models with subsystem 
symmetry.  In the gauge-fixed Hamiltonian (\ref{eq:H_gauge_fixed}), $A_{i}$ then becomes the familiar 
$\mathbb{Z}_{2}$ charge operator, while the operator $B_{n}$, appearing in the effective Hamiltonian 
$H_{\mathrm{eff}}$, precisely measures the $\mathbb{Z}_{2}$ flux through an elementary plaquette. 

\begin{table*}
\begin{tabular}{|>\centering{c}|>\centering{c}|L|}
  \hline
$\boldsymbol{H_{\mathrm{spin}}}$
&$\boldsymbol{H_{\mathrm{nexus}}}$ &  \begin{centering}{\bf Explanation}\end{centering} \,\,\\
    \hline
 $\mathcal{O}^{(a)}_{i}$ & $\sigma^{z}_{i,a}$ & 
 \begin{flushleft} A classical configuration of Ising spins ($\tau$) may be specified by spins ($\sigma$) placed at the centers of each of the interaction terms $\mathcal{O}_{i}^{(a)}$ for the Ising matter. 
 \end{flushleft}\\
 \hline
   $\tau^{x}_{i}$ & $A_{i}$ &
   \begin{flushleft} The dual representation of $\tau^{x}_{i}$ is given by the nexus charge $A_{i}$, defined as the product of $\sigma^{x}_{i}$ terms that correspond to the interactions flipped by $\tau^{x}_{i}$.   \end{flushleft}  \\
 \hline 
 \hline
 $\displaystyle\prod_{(i,a)\in Q(j)}\mathcal{O}^{(a)}_{i}=1$ & $B_{j}^{(1)}\ket{\Psi} = \ket{\Psi}$ & 
 \begin{flushleft} A local product of interactions for the Ising matter fields that yields the identity corresponds to a constraint on the dual Hilbert space. The constraint restricts the Hilbert space of $\{\sigma\}$ to that of \emph{domain wall} configurations in the ordered phase of $H_{0}$. \end{flushleft}\\
 \hline
 $\widetilde{W} \equiv \displaystyle\prod_{(i,a)\in \Sigma}\mathcal{O}^{(a)}_{i}$ & ${W} \equiv \displaystyle\prod_{(i,a)\in \Sigma}\sigma^{z}_{i,a}$ & 
 \begin{flushleft} A product of $\sigma^{z}$ operators is dual to a product of interaction terms for the matter fields. As a result, the nexus charge is a fracton only if there is no product of interaction terms that can create an isolated pair of spin-flips.  \end{flushleft}\\
 \hline
   \end{tabular}
 \caption{Correspondence between operators in the Hilbert spaces of the Ising matter fields and the nexus spins, obtained from the \emph{F-S} duality.}  
 \label{fig:Duality_Table}
\end{table*}

Within our construction, the $B_{i}^{(k)}$ operators provide the natural generalization of the flux in a lattice gauge theory.  As the excitation obtained by flipping the eigenvalue of a $B_{i}^{(k)}$ operator 
is often point-like, we will refer to the excitation as a ``generalized monopole".  While the flux is always a 
line-like excitation in a three-dimensional Abelian lattice gauge theory, the behavior of the ``generalized 
monopole" can be quite varied.    As an example, the generalized monopole is a fracton in both the 
checkerboard spin model and in Haah's code, but is free to move along a line without creating additional 
excitations in the X-cube model.  We refer to such an excitation as a dimension-1 quasiparticle \cite{Fracton}, as the excitation is only mobile along a line.  As shown in Figure 
\ref{fig:X_Cube}b, a straight Wilson line can create an isolated pair of the generalized monopole excitations 
in the X-cube model, which can only move along the line without creating additional excitations. Within 
type-I fracton topological order, composites of the fracton charge excitations that are mobile in two 
dimensions can have non-trivial mutual statistics with the generalized monopole.  This is true in both the 
checkerboard and X-cube models, where an anyon formed from a composite of two fracton charges has 
$\pi$-statistics with a generalized monopole in its plane of motion.    

\begin{figure}
\includegraphics[trim = 0 0 0 0, clip = true, width=0.34\textwidth, angle = 0.]{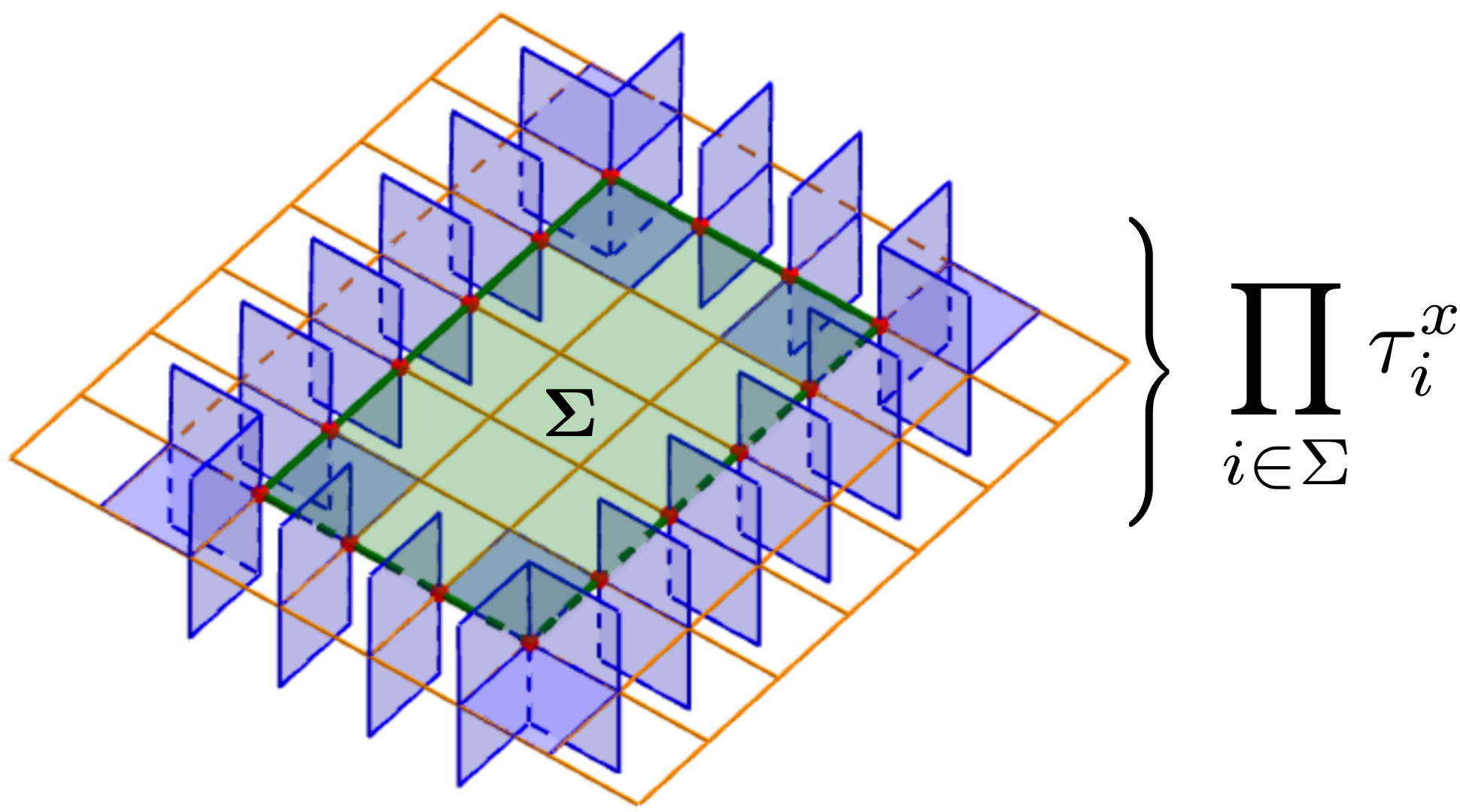}
\caption{A ``domain wall" in the ground-state of the plaquette Ising model ($H_{\mathrm{plaq}}$) is depicted, 
by coloring the plaquette interactions that have been flipped by the action of a spin-flip transformation along 
a planar region $\Sigma$. 
The \emph{F-S} duality implies that the ground-state for the X-code fracton phase is given by an equal 
superposition of a dual representation of these domain-walls.  }
  \label{fig:Plaquette_Domain_Wall}
\end{figure}

We now consider the generalized lattice gauge theory (\ref{eq:H_gauge_fixed}) when $J = 0$, so that the nexus field (defined in the $\sigma^{z}$-basis) has no dynamics. The resulting Hamiltonian
\begin{align}\label{eq:H_nexus}
H_{\mathrm{nexus}} = -t\sum_{i,a}\sigma^{z}_{i,a} -h\sum_{i}A_{i}
\end{align}
has a local symmetry, as the $B_{i}^{(k)}$ operators commute with each term in (\ref{eq:H_nexus}). We refer 
to the emergent local constraints on the Hilbert space 
\begin{align}\label{eq:Flatness}
B_{i}^{(k)}\ket{\Psi} = \ket{\Psi}
\end{align}
as the generalized ``flatness" condition, analogous to a flat connection in a continuum gauge theory, as these 
constraints are obtained in the limit that there is zero ``flux" of the nexus field.

Our construction of a generalized lattice gauge theory implies that the quantum dual of the Ising matter 
in the presence of a transverse field:
\begin{align}\label{eq:Spin}
H_{\mathrm{spin}} = -t\sum_{i,a}\mathcal{O}_{i}^{(a)}[\tau^{z}] - h\sum_{i}\tau^{x}_{i}.
\end{align}
is precisely given by the nexus Hamiltonian 
(\ref{eq:H_nexus}), combined with the generalized flatness condition (\ref{eq:Flatness}). 

Without appealing to the generalized lattice gauge theory, the duality can be obtained directly from the Hamiltonian $H_{\mathrm{spin}}$. A dual representation is constructed by 
placing the nexus spins at the centers of the interactions $\mathcal{O}_{i}^{(a)}$.  The nexus spins are now 
interpreted as domain wall variables for the ordered phase ($t/h\gg 1$) of the spin model 
$H_{\mathrm{spin}}$.   The nexus spins must satisfy local constraints due to the geometry of the multi-spin 
interactions $\mathcal{O}_{i}^{(a)}[\tau]$ in order to correspond to the physical space of domain walls 
between ground-states of $H_{0}$. These local constraints are precisely given by the generalized flatness 
condition (\ref{eq:Flatness}).  As an example, the generalized monopole operators 
$B_{i}^{(k)}$ for the plaquette Ising model are obtained by noting that the product of four plaquette 
interactions that wrap a cube is equal to the identity.  Our \emph{F-S} duality implies a map between local 
operators in the Hilbert spaces of the Ising matter and nexus fields, as summarized in Table 
\ref{fig:Duality_Table}. As an example, a domain wall in the ground-state of the plaquette Ising model is 
shown in Figure \ref{fig:Plaquette_Domain_Wall}.  

\section{Fracton Topological Order} 

We now invoke the \emph{F-S} duality to demonstrate that the commuting Hamiltonian 
\begin{align}\label{eq:H_fracton}
H_{\mathrm{fracton}} = -\sum_{i,k}B_{i}^{(k)} - \sum_{i}A_{i}
\end{align}
exhibits fracton topological order. We argue that (i) the spectrum of $H_{\mathrm{fracton}}$ has 
sub-extensive topological degeneracy and that (ii) the nexus charge is fundamentally immobile.  We provide 
rigorous proofs of these statements in the Appendix using techniques in commutative algebra and 
elementary algebraic geometry, which provide effective mathematical tools to study the 
subsystem 
symmetries of classical 
spin models, as well as the ground-state degeneracy and excitation spectrum of fracton topological phases.  
An algebraic representation of a classical Ising system defines an algebraic variety over the field of 
characteristic 2 ($\mathbb{F}_{2}$), defined by $\mathbb{Z}_{2}$ addition and multiplication \cite{Polynomial}.
Two conditions on this variety, as derived in the Appendix from the Buchsbaum-Eisenbud criterion 
\cite{Eisenbud, Buchsbaum_Eisenbud} for the exactness of a 
complex of free modules, guarantee that the quantum dual exhibits fracton topological order. 

We begin by using the \emph{F-S} duality to demonstrate that 
the sub-extensive degeneracy of the classical, $h=0$ ground-state of 
$H_{\mathrm{spin}}$ implies that the Hamiltonian $H_{\mathrm{fracton}}$ has sub-extensive 
\emph{topological} 
ground-state degeneracy on the torus.  Recall that a product of $\tau^{x}_{i}$ operators along an appropriate 
subsystem $\Sigma$ generates a
symmetry of the Hamiltonian $H_{\mathrm{spin}}$.  When $t/h\gg 1$, the ground-state exhibits classical, 
sub-extensive degeneracy since 
there are $O(L)$ independent subsystems along which a spin-flip commutes with all of the interaction terms 
$\mathcal{O}_{i}^{(a)}$.  The plaquette Ising model, for example, commutes with the product of $\tau_{i}^{x}
$ along a plane, and the ground-state has sub-extensive, classical degeneracy since 
there are $O(L)$ independent planes along which a spin-flip may be performed.

From the operator dictionary for the \emph{F-S} duality, the dual representation of this spin-flip operator is 
given by a product of nexus charges $A_{i}$ along the same subsystem $\Sigma$.  Furthermore, each 
interaction term $\mathcal{O}_{i}^{(a)}$ is dual to a single-spin operator $\sigma^{z}_{i,a}$.  Since the 
\emph{F-S} duality preserves the commutation relations between operators, we conclude that due to the 
subsystem symmetry of $H_{\mathrm{spin}}$, the operators in the dual theory satisfy:
\begin{align}
\left[\sigma^{z}_{i,a},\,\,\prod_{i\in\Sigma}A_{i}\right] = 0
\end{align}
for all $i$, $a$.  This commutation relation can only be satisfied if the product of nexus charges along $
\Sigma$ yields the identity, so that $\prod_{i\in\Sigma}A_{i} = 1$.  
This relation implies that not all of the nexus charge operators are independent on the torus. Each of the $O(L)$ independent 
subsystems associated with the subsystem symmetry of $H_{\mathrm{spin}}$ reduces the number of independent nexus charge operators by one.  We denote the total number of such independent subsystems by $k_{A}$.

In the ground-state of the fracton Hamiltonian, the  $2^{N}$-dimensional Hilbert space of $N$ nexus spins is constrained by the $M$ nexus charge and monopole operators that appear in the Hamiltonian.  However,   only $M - k$ of the operators are independent on the torus, where $k = k_{A} + k_{B}$ is the number of ``dependency relations" on both the nexus charge and monopole operators. The topological ground-state degeneracy on the torus is given by $D = 2^{k + (N-M)}$.
When the number of interactions appearing in 
$H_{\mathrm{fracton}}$ is identical to the total number of 
nexus spins ($N=M$),
as is the case for all of the fracton models considered in this work, the topological 
degeneracy is precisely $D = 2^{k}$. 
In this case, the sub-extensive degeneracy of the $h=0$ ground-state 
of the spin model $H_{\mathrm{spin}}$ provides a lower bound on the 
topological degeneracy of $H_{\mathrm{fracton}}$. 
For example, the checkerboard spin model has topological ground-state degeneracy $\log_{2}D = 6L-6$ on the length-$L$ three-torus, as we compute in the Appendix, while the tetrahedral Ising model only has classical degeneracy $\log_{2}D_{c} \sim O(3L)$ since the model has subystem symmetries along three orthogonal planes.

In addition to the sub-extensive, topological degeneracy of $H_{\mathrm{fracton}}$, we also wish to show 
that that there is no degeneracy in the spectrum of 
Hamiltonian due to the presence of \emph{local} observables. In the absence of local observables, the local reduced density matrix will be identical for any degenerate states in the spectrum of the Hamiltonian, and the topological degeneracy, as computed by constraint-counting, will be stable to local perturbations \cite{BHM}.
As we demonstrate in the Appendix, the
ground-states of $H_{\mathrm{fracton}}$ are guaranteed to be locally indistinguishable, 
provided that the classical spin system $H_{0}$ has no lower-dimensional symmetries along subsystems of 
dimension $d_{s} < 2$ (e.g. line-like symmetries).   We prove this by using an algebraic 
representation of $H_{\mathrm{fracton}}$ and also argue this as a consequence of the \emph{F-S} 
duality.  

Having demonstrated that $H_{\mathrm{fracton}}$ exhibits sub-extensive topological degeneracy and that the 
degenerate ground-states are locally indistinguishable, we now demonstrate that the nexus charge is 
indeed a fracton excitation, provided that the spin model 
(\ref{eq:Spin}) satisfies a simple condition. Consider acting on the ground-state of $H_{\mathrm{fracton}}$ 
with the operator 
\begin{align}
W \equiv \prod_{(i,a)\in\Sigma}\sigma^{z}_{i,a}
\end{align}
where $\Sigma$ is some subset of the lattice.  The operator $W$ will create nexus 
charge excitations by anti-commuting with a collection of $A_{i}$ operators. Invoking the \emph{F-S} duality, 
we observe that the pattern of excitations created by $W$ is precisely given by the location of spin-flips 
created by the dual operator $\widetilde{W} \equiv \prod_{(i,a)\in\Sigma}\mathcal{O}_{i}^{(a)}[\tau^{z}]$ when 
acting on the paramagnetic state $\ket{{\Psi}_{\mathrm{para}}} \equiv \ket{\rightarrow\cdots\rightarrow}$.  
The spectrum of $H_{\mathrm{fracton}}$ contains fractons only if there is \emph{no operator} of the form 
$\widetilde{W}$ that can create a single pair of spin-flip excitations.  If such an operator did exist, then it 
would be possible to move a single nexus charge without any energy cost and the charge would be mobile.

\begin{figure}
$\begin{array}{c}
  \includegraphics[trim = 3 0 4 4, clip = true, width=0.47\textwidth, angle = 0.]{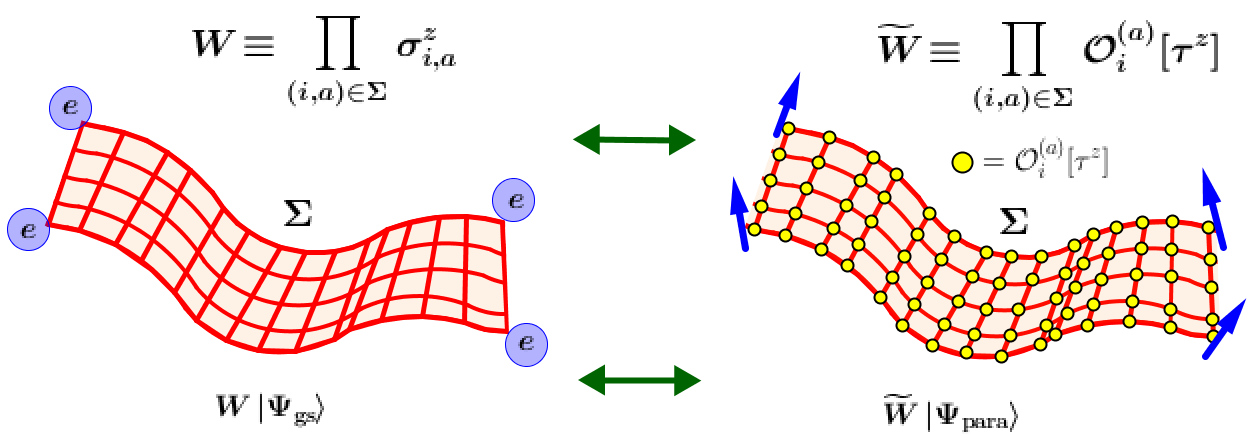}\\  \text{(a)} \hspace{1.5in} \text{(b)}\\\\\\
\hspace{.1in} \includegraphics[trim = 3 0 4 4, clip = true, width=0.47\textwidth, angle = 0.]{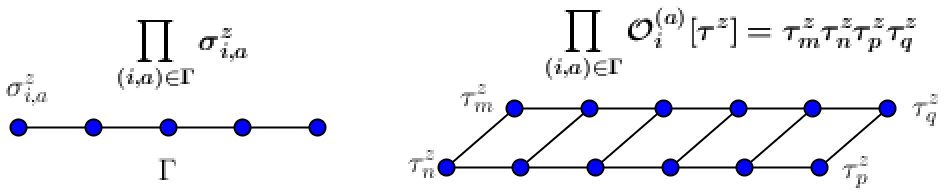} \\ 
\text{(c)} \hspace{1.5in} \text{(d)}
 \end{array}$
 \caption{The nexus charge is a fracton only if there is no operator $W$ that can create an isolated pair of 
 excitations when acting on the ground-state of $H_{\mathrm{fracton}}$, as in (a).  Equivalently, a dual 
 representation of the operator, given as a product of the interaction terms in the quantum dual as shown in 
 (b), cannot create an isolated pair of spin-flips when acting on the paramagnetic state 
 $\ket{\Psi_{\mathrm{para}}} \equiv \ket{\rightarrow\cdots\rightarrow}$. An example is given in (c) and (d); a 
 straight Wilson line acting on the ground-state of $H_{\mathrm{X}\text{-}\mathrm{Cube}}$ in (c) admits a 
 dual representation as a product of four-spin plaquette interactions along a line, as shown in (d).  No 
 product of interaction terms in the plaquette Ising model can produce an isolated pair of spin-flips.  As a 
 result, the nexus charge in $H_{\mathrm{X}\text{-}\mathrm{Cube}}$ must be a fracton.}
  \label{fig:Duality_Fig}
\end{figure}

Our condition for the existence of fracton excitations is simple to demonstrate for the plaquette Ising and 
tetrahedral Ising Hamiltonians.  Here, it is evident that any product of the four-spin interactions shown in 
Table \ref{fig:Classical_Spins} creates \emph{at least} four spin-flip excitations 
when acting on the paramagnetic state $\ket{\widetilde{\Psi}}$.  Therefore, the nexus charge for each of the 
corresponding $H_{\mathrm{fracton}}$ Hamiltonians is a fracton.  In fact, if any Ising Hamiltonian $H_{0}$ 
has a subsystem symmetry along three orthogonal planes, 
and two or more independent interactions per lattice site, then the quantum dual will always exhibit 
fracton 
topological order, as we demonstrate in the Appendix.  All Type-I fracton topological phases that have been 
discovered thus far fit into this framework.

We now summarize the precise conditions on the classical spin system $H_{0}$, as derived in the Appendix using the algebraic representation of the classical spin system, that guarantee that $H_{\mathrm{fracton}}$ exhibits fracton topological order:
\begin{enumerate}
\item $H_{\mathrm{spin}}$ contains more than one independent interaction term per lattice site
\item No product of the interaction terms ($\mathcal{O}_{i}^{(a)}$) can generate an isolated pair of spin-flips 
\end{enumerate}

\section{Phase Diagram} 

We discuss the phase diagram of the spin-nexus Hamiltonian (\ref{eq:H_ss}).
In the limit $t \ll h$,
the nexus field decouples from the Ising matter;
$\tau_i^x$ is set to 1, and the effective Hamiltonian for the nexus field is
\begin{align}
H_{ t \ll h} = - \tilde t \sum_{i,k} B_i^{(k)} - J \sum_i \sigma^x_i.
\end{align}
The generalized Gauss' law becomes $A_i = 1$.
Here, $\tilde t$ is some power of $t$ as $B_i^{(k)}$
is obtained from perturbation theory.
When $J \ll \tilde t$,
the nexus field forms a fracton phase that is described by Hamiltonian $H_{\mathrm{fracton}}$.
The topologically-ordered fracton phase and the Ising paramagnet survive up to a finite $t/h$ and $J/h$,
as both phases are gapped and stable to perturbations \cite{BHM}.

\emph{Confinement:} 
From the topologically-ordered fracton phase, we may proceed in two directions.  First, we consider increasing $J/h$ while keeping $t/h \ll 1$, a constant.  Above a critical value $(J/h) \ge (J/h)_{*}$, the ground-state will be a condensate of  nexus flux excitations and the fracton topological order will be destroyed.  The nature of the transition between the fracton phase and the trivial (confined) phase is currently unknown.

\begin{figure}
$\begin{array}{cc}
  \includegraphics[trim = 0 0 0 0, clip = true, width=0.22\textwidth, angle = 0.]{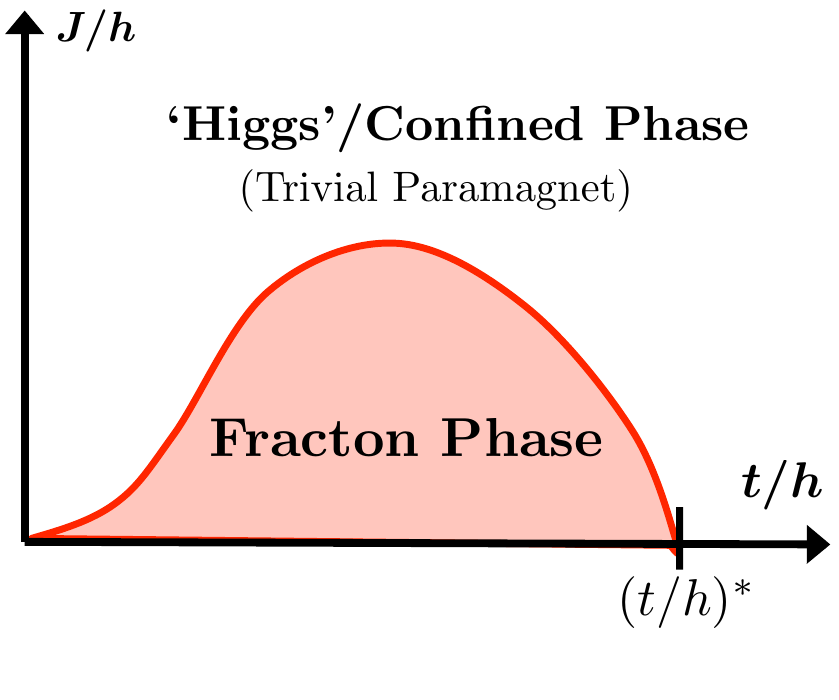} &
 \includegraphics[trim = 5 61 135 23, clip = true, width=0.25\textwidth, angle = 0.]{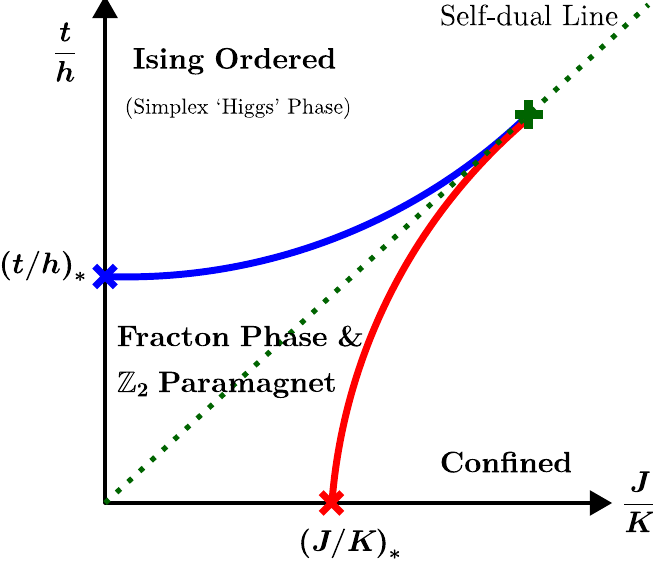}\\
 \text{(a)} & \text{(b)}
 \end{array}$
 \caption{Schematic phase diagram of (a) the spin-nexus Hamiltonian (\ref{eq:H_ss}). The `Higgs' phase for the nexus field is smoothly connected to the phase reached by condensing the nexus charge.  The checkerboard model coupled to Ising matter fields admits an additional \emph{self-duality} under the exchange of the nexus charge and flux; as a result, the phase diagram is as shown in (b).}
  \label{fig:Phase_Diagram}
\end{figure}

\emph{`Higgs' Phase:} 
We now consider the region of the phase diagram where $t \gg h$,
keeping $J \ll h$ at a fixed constant.
Here, the matter fields enter an ordered state 
with $\langle \sigma^z_i \mathcal{O}_{i}^{(a)}[\tau^{z}]\rangle = 1$.
This may be seen as the analog of a `Higgs' phase,
as the Ising order gives the nexus field a mass $m\sim O(t)$ that destroys the fracton topological phase.
The ground-state in this region of the phase diagram is {non-degenerate},
even though the ordered phase of the pure spin model (\ref{eq:Spin})
has sub-extensive degeneracy.
We may demonstrate this by observing that in the gauge-fixed Hamiltonian (\ref{eq:H_gauge_fixed}),
increasing $t/h$ destroys the fracton topological order 
by condensing the nexus charge,
and produces a non-degenerate ground-state.
We also observe from $H_{\mathrm{eff}}$ that the confined and Higgs regions of the spin-nexus phase diagram
are smoothly connected,
as in the Ising lattice gauge theory.
We summarize our schematic phase diagram in Fig. \ref{fig:Phase_Diagram}a.

In passing, we observe that the checkerboard fracton Hamiltonian in Table~\ref{fig:Classical_Spins}
in the presence of two transverse fields
$H = - K \sum_{c} B_{c} - h\sum_{c}A_{c} - \sum_{i,j}(J\,\sigma^{x}_{i,a} + t\,\sigma^{z}_{i,a})$
has a symmetry  under $\sigma^z \Leftrightarrow \sigma^x$.
This implies that the phase diagram should be symmetric under $K \leftrightarrow h$ and $J \leftrightarrow t$.
The confinement and `Higgs' transition must be dual to each other, and
the line of phase transitions must meet at a self-dual point of the phase diagram.
It is unknown whether any of the indicated phase transitions in Fig. \ref{fig:Phase_Diagram} are continuous.

\section{Concluding remarks}

Translationally-invariant, commuting Hamiltonians built from interacting qubits \cite{Polynomial}
and fermions \cite{Fracton}
admit a convenient algebraic representation as a collection of polynomials over a finite field.
The translation group of the lattice is $\mathbb Z^D$, whose group algebra happens to be the polynomial algebra.
The polynomials conveniently keep track of the support of various operators.
Remarkably, this algebraic characterization of the Hamiltonian terms
enable us to decide whether the given Hamiltonians is commuting, degenerate, topologically ordered,
and the nature of the excitations \cite{Fracton, Polynomial}.
Also, it gives a unique method to calculate the ground-state degeneracy of our exotic models.
In the context of our nexus theory,
the polynomial representation has a natural physical interpretation,
as it precisely specifies the generalized Gauss' law $G_{n}$
that defines the spin-nexus Hamiltonian (\ref{eq:H_ss}).
In this way, the polynomial representation
encodes the local symmetry that defines a fracton topological phase.
We elaborate on these method in the appendix, which we aim to be pedagogical.

With the identification of a generalized Gauss's law that characterizes a fracton topological phase,
our work provides an important step towards searching for material realizations of fracton topological order.
Such a local conservation law can, in principle, appear in physical systems such as frustrated magnets,
where our generalized gauge theory can emerge
as an effective description at low energies, leading to fracton topological order.

\emph{Note Added:} Near the completion of this work, we were informed that a related work is being written up \cite{Dom_Private_Comm}. 

\acknowledgments
This work was supported by the David and Lucile Packard Foundation (LF and SV) and the Pappalardo Fellowship in Physics at MIT (JH).

\appendix

\section{Algebraic Construction of a Fracton Hamiltonian from a Classical Spin System}

Commutative algebra and elementary algebraic geometry 
provide an indispensable machinery for demonstrating the results summarized in the main text.
Starting from a purely algebraic representation of commuting Pauli Hamiltonians,
as originally introduced in Ref. \cite{Polynomial_Bib},
we demonstrate that certain classical spin systems with subsystem symmetry
may be used to construct a commuting Hamiltonian with fracton topological order.
Only the data contained in the classical spin model
is required as input to generate the topologically-ordered fracton Hamiltonian.
This map between a classical spin system and a topologically-ordered Hamiltonian
may be used to construct a wide variety of topological orders,
ranging from conventional $\mathbb{Z}_{N}$ {gauge theory}
to the type-I and type-II fracton topological orders presented in the main text.
Our presentation of this correspondence in this section,
while mathematically inclined,
is self-contained and includes a rigorous and precise statement
of the results summarized in the main text.

Given a translationally-invariant Pauli Hamiltonian 
(also called stabilizer or additive code Hamiltonian)
it is possible to write a purely algebraic representation of the Hamiltonian,
known as the {stabilizer map} \cite{Polynomial_Bib}.
For a Hamiltonian defined on a $d$-dimensional lattice with an $\ell$-site basis,
and containing $m$ distinct operators per lattice site,
the stabilizer map will be a {$2\ell \times m$}
matrix of Laurent polynomials over the finite field $\mathbb{F}_{2}$.
{The prescription to obtain the stabilizer matrix is as follows.
Given a Pauli operator (a tensor product of Pauli matrices),
write it as a product of two Pauli operators,
where one entirely consists of $\sigma^x$ and the other of $\sigma^z$, using $\sigma^y = i \sigma^x \sigma^z$.
Select a reference site $s_0$. Relative to $s_0$, an arbitrary site where a Pauli matrix acts on
will be given by an integer displacement
vector, e.g., $(a,b,c)$ in three dimensions, and one writes it as $x^a y^b z^c$.
Thus, every nontrivial factor $\sigma^x$ or $\sigma^z$ gives a monomial $x^a y^b z^c$.
Collect all monomials corresponding to $\sigma^z$ that act on $a$-th spins in the unit cell,
and combine them with $+$. This give a Laurent polynomial that goes into $a$-th component in a column vector.
The sum of monomials corresponding to $\sigma^x$ goes into $(a+\ell)$-th component.
This way, one obtains a column vector of Laurent polynomials associated with the Pauli operator.
For each type of term in the Hamiltonian, repeat the process to obtain a collection of column vectors, that is a matrix.
Note that the overall sign of the Pauli operator is dismissed.
We denote by $p(O)$ the Laurent polynomial vector corresponding to the Pauli operator $O$.
}

From the stabilizer map $S$, 
we may then {define} the \emph{excitation map} {$E$}, 
defined so that $E\cdot p(\mathcal{O})$ yields the pattern of excitations
created by $\mathcal{O}$ when acting on the ground-state of the stabilizer Hamiltonian.
As {shown} in \cite{Polynomial_Bib},
the excitation map {is given by} $E \equiv S^{\dagger}\lambda_{\ell}$
with $S^{\dagger} \equiv (\overline{S})^{T}$ {and} {where}
\begin{align}
\lambda_{\ell} \equiv \left(\begin{array}{cc} 0 & \boldsymbol{1}_{\ell\times \ell}\\
\boldsymbol{1}_{\ell\times \ell} & 0 \end{array}\right).
\end{align}
Here, $\overline{S}$ indicates that each {monomial} in the stabilizer map
has been inverted,
e.g. for the polynomial $f(x,y,z)$ we {have}
$\overline{f(x,y,z)} = f(\overline{x}, \overline{y}, \overline{z})$ 
with $\overline{x} \equiv x^{-1}$, $\overline{y} \equiv y^{-1}$, $\overline{z} \equiv z^{-1}$.
The following are necessary conditions for the stabilizer map to correspond to a topologically-ordered stabilizer Hamiltonian.
First, all terms in the Hamiltonian commute 
if and only if $S^{\dagger}\lambda_{\ell}S = E\cdot S = 0$ \cite{Polynomial_Bib}.
Second, we require that any degenerate ground-states on the torus are locally indistinguishable,
so that the degeneracy is topological in nature.
This is guaranteed if, on an infinite system, {$\ker E = \mathrm{im}~S$.}
{This means that}
any local operator commuting with the stabilizer Hamiltonian
is always a product of operators already appearing in the Hamiltonian.

Consider the classical spin Hamiltonian $H_{0}$,
defined on a Bravais lattice {with a single spin per lattice site}.
The stabilizer map for this classical spin Hamiltonian takes the form
\begin{align}
S_{c} \equiv \left(\begin{array}{cccc}
f_{1} & f_{2} & \cdots & f_{n}\\
0 & 0 & \cdots & 0
\end{array}\right)
\end{align}
As the Hamiltonian is classical and contains only $\tau^{z}$-type terms,
the second row in the stabilizer map is identically zero.

{
As an example, consider the plaquette Ising model.
There are three plaquette terms depending on their orientation.
The plaquette in the $xy$-plane couples spins at
$(0,0,0)$, $(1,0,0)$, $(0,1,0)$, and $(1,1,0)$
by $\tau^z$ operators.
The position coordiantes are translated into the monomials as $1$, $x$, $y$, $xy$, respectively,
and thus the the plaquette term is mapped to $1+x+y+xy$.
Likewise, the plaquette term in $yz$-plane is represented by $1+y+z+yz$,
and that in $zx$-plane is by $1+x+z+xz$.
The stabilizer map is
}
\begin{align}
S_\text{plaq} = 
\begin{pmatrix}
1+x+y+xy & 1+y+z+yz & 1+x+z+xz \\ 0 & 0 & 0
\end{pmatrix}.
\label{eq:IsingPlaquetteStabilizerMap}
\end{align}
{Below, we will often omit the zero rows, 
which will not cause any confusion when it is clear from the context
whether the operator is of $\sigma^z$ or $\sigma^x$ type.}

We now 
{summarize our main results using algebraic language.}
Given the stabilizer map of the form $S_{c}$
for a classical spin Hamiltonian in $d$ spatial dimensions,
consider the \emph{ideal} $I(S_{c}) \equiv \langle f_{1}, \ldots, f_{n}\rangle$,
defined as the set of all linear combinations of the polynomials $f_{1}$, \ldots, $f_{n}$ in the stabilizer map:
\begin{align}
I(S_{c}) \equiv \left\{\sum_{i{ = 1}}^{ n}c_{i}f_{i}\,\,\Big|\,\,c_{i}\in\mathbb{F}_{2}[x_{1}^{\pm}, \ldots, x_{d}^{\pm}] \right\}.
\end{align}
If the ideal $I(S_{c})$ constructed from the stabilizer map
satisfies the following conditions,
then 
{the quantum dual obtained by our prescription in the main text exhibits fracton topological order.}
\begin{enumerate}[label=\bfseries \arabic*.]
\item \textbf{Co-dimension Condition:} The stabilizer ideal must have a sufficiently large co-dimension {(height)}
\begin{align}
\mathrm{codim}\,I(S_{c})\, \ge 2.
\end{align}
Physically, this means that the classical spin Hamiltonian must have 
at least {two} independent interactions per lattice site.
This condition alone ensures 
that the quantum dual of the classical model with transversal field 
is topologically ordered.

\item \textbf{Fracton Condition:} The stabilizer ideal $I(S_{c}) \equiv \langle f_{1}, \ldots, f_{n}\rangle$ 
{does not} contain binomial terms,
\begin{align}
1 + x_{1}^{n_{1}}\cdots x_{d}^{n_{d}} \notin I(S_{c})
\end{align}
for all $n_{1},\ldots, n_{d}> 0$.
Physically, this guarantees that, in the topologically-ordered Hamiltonian
built from the algebraic data in the classical spin model, 
{the elementary topological excitations are immobile.}

\item \textbf{Planar symmetries imply the fracton condition:}
Planar symmetries on $xy$, $yz$, and $zx$ planes imply 
$f_i(1,1,z) = 0$, $f_i(x,1,1)=0$, and $f_i(1,y,1) = 0$ for all $i$, respectively.
The latter conditions combined imply the fracton condition.
\end{enumerate}

If the first two conditions are satisfied, 
then we may build a commuting Hamiltonian 
with fracton topological order from the data in the classical stabilizer map $S_{c}$.
The topological ground-state degeneracy of this Hamiltonian
will take the general form 
{$\log_{2}D \sim c_{1}L^{d-2}$ }
on the length-$L$ torus.

\subsection{Symmetries}\label{sec:symmetry}

An operator $\mathcal{O}$,
given by a product of $\tau^{x}$ terms,
generates a subsystem symmetry
if $\mathcal{O}$ commutes with the classical Hamiltonian $H_{0}$.
In the polynomial representation,
the non-local operator $\mathcal{O}$ is expressed by a formal infinite sum 
$h(O) = \sum_{\vec s} x^{s_1} y^{s_2} z^{s_3}$
where $\vec s$ ranges over all spins where $\mathcal O$ acts on by $\tau^x$.
The commutativity of $\mathcal O$ with the Hamiltonian is equivalent to
the requirement that $\mathcal O$ does not create any excitations when acting on the ground-state.
This is the case if and only if 
the image of $h(\mathcal O)$ under the excitation map $E = S^\dagger \lambda_1$
is zero: $E\, h(\mathcal O) = 0$, or equivalently, $h(\bar x,\bar y,\bar z) S = 0$.

As an example, consider an operator $O_{xy}$ that flips all spins on an $xy$-plane.
The polynomial representation reads
\begin{align}
h_{xy} = \sum_{n,m \in \mathbb Z} x^{n}y^{m} = \overline{h_{xy}}.
\end{align}
This formal infinite series has a property that $x h_{xy} = h_{xy} = y h_{xy}$,
which is just another expression of the fact that $O_{xy}$
is translation invariant within the $xy$-plane.
Consequently, for any Laurent polynomial $f(x,y,z)$ we have $f(x,y,z) h_{xy} = f(1,1,z) h_{xy}$.
We observe that this is a symmetry of the plaquette Ising model,
whose stabilizer map is found in \eqref{eq:IsingPlaquetteStabilizerMap},
since 
\begin{align*}
h_{xy} \cdot (1+x+y+xy) &= h_{xy}\cdot (1+1+1+1) = 0 ,\\
h_{xy} \cdot (1+y+z+yz) &= h_{xy} \cdot (1+1 + z + z )  =0,\\
h_{xy} \cdot (1+x +z +xz)& = h_{xy} \cdot (1 + 1 + z + z ) =0.
\end{align*}
Due to the translation invariance of the model,
the operator $O_{xy}$ at any $xy$-plane is a symmetry.
In the polynomial representation, the operator $O_{xy}$ at $z=c$-plane
is expressed as $z^c h_{xy}$ and the above equation is trivially satisfied.
We leave it to readers to verify that similar operators on $yz$- and $zx$-planes
are symmetries of the plaquette Ising model.

\subsection{F-S duality}

Recall that the local symmetry operator $G_i = \tau^x_i A_i$
was defined by $A_i = \prod_{a \in P(i)} \sigma^x_{i,a}$
where $P(i)$ is the set of all classical interaction terms $\mathcal O_i^{(a)}$
that anti-commutes with $\tau^x_i$.
In the polynomial formulation, $P(i)$ is exactly the image of $p(\tau^x_i)$
(the polynomials representing $\tau^x_i$, which is a unit vector)
under the excitation map $S_c^\dagger \lambda_1$.
This means that the polynomial representation of the operator $A_i$
is the conjugate transpose of the first row of $S_c$.

The ``flatness'' constraint operator $B_k$
is obtained by considering the nontrivial local product of $\mathcal O_i^{(a)}$
that becomes the identity.
Since the identity is a zero vector in the polynomial representation,
Each $B_k$ operator corresponds to a nontrivial relation
\begin{align}
\sum_{a=1}^\ell b_a p(O_i^{(a)}) = 0
\end{align}
where $p(\mathcal O_i^{(a)}) = (S_c)_a$ is the polynomial representing the $\tau^z$-type
classical interaction term.
The collection of vectors $b$ in the relation is the kernel of the matrix $S_c$.
There are finitely many {\em generators} $G^{(k)}$ of $\ker S_c$,
where $G^{(k)} \in \mathbb{F}_{2}[x^\pm,y^\pm,z^\pm]^{\ell}$.
(The fact that there are only finitely many generators 
is a property of the ring $\FF_2[x^\pm, y^\pm, z^\pm]$ being Noetherian.)
These generators are nothing but the polynomial representation of $B_k$.

\begin{figure}
 \includegraphics[trim = 100 145 17 124, clip = true, width=0.34\textwidth, angle = 0.]{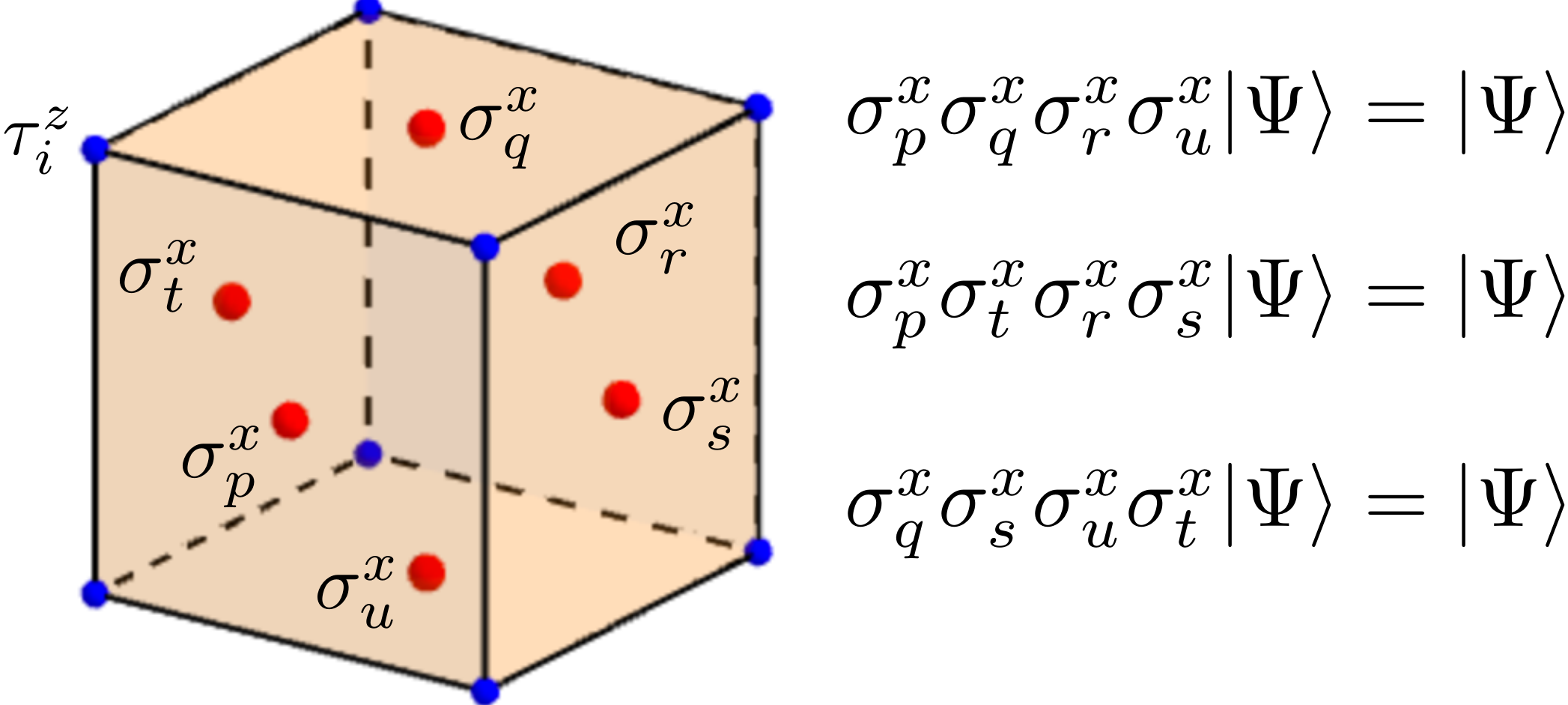}
 \caption{Dual representation of the plaquette Ising model in the presence of a transverse field.  We place nexus spins at the center of each four-spin plaquette interaction, so that $\sigma^{x}_{p}\equiv \prod_{i\in\partial p}\tau^{z}_{i}$. The product of four adjacent four-spin interactions that wrap around the cube is equal to the identity (e.g. the product of plaquette interactions $p$, $q$, $r$ and $u$).  In the dual representation, this leads to the indicated constraints at each cube.  Only two of the three constraints are independent.}
  \label{fig:X-Cube_Appendix}
\end{figure}

For the plaquette Ising model,
$\ker S_{\mathrm{plaq}}$ has two generators 
\begin{align}
G^{(1)} &= (1 + z, 1 + x, 0)^{T}\\
G^{(2)} &= (0, 1 + x, 1 + y)^{T}.
\end{align}
Each generator encodes the product of the appropriate plaquette interactions
at each cube that yield the identity,
as shown in Fig.~\ref{fig:X-Cube_Appendix}.

Summarizing, given a classical Hamiltonian represented by a row matrix $S_c$ of Laurent polynomials,
we have obtained a quantum model whose Hamiltonian terms
corresponds to the columns of the following stabilizer map.
\begin{align}
\boxed{S_{f} = S_c^\dagger \oplus G = \begin{pmatrix}S_c^\dagger & 0 \\ 0 & G \end{pmatrix}}
\end{align}
where
\begin{align}
G = \begin{pmatrix} G^{(1)} & \cdots & G^{(m)} \end{pmatrix}
\end{align}
is a matrix enumerating generators of $\ker S_c$ in its columns.
The excitation map is then
\begin{align}\label{eq:E_f_map}
E_{f} = S_f^\dagger \lambda_\ell =
\left(\begin{array}{cc} 
0 & S_{c}\\
G^{\dagger} & 0
\end{array}\right).
\end{align}
Physically, this Hamiltonian is defined on a Bravais lattice with an $\ell$-site basis.
At each lattice site, the Hamiltonian has a single ``$\sigma^{x}$-type" (nexus charge) interaction
and $m$ ``$\sigma^{z}$-type" (nexus flux) interactions.

A wide range of conventional and fracton topological orders
are described by stabilizer maps of the form $S_{f} = S_c^\dagger \oplus G$.
Physically, this means that these topologically ordered states
are built by ``gauging" the symmetry of a classical spin system,
which can be easily read off from the stabilizer map $S_{f}$
for the topologically ordered state.

First, the stabilizer map built from the plaquette Ising model (see \eqref{eq:IsingPlaquetteStabilizerMap})
is given as 
\begin{align}
S_{\mathrm{X}\text{-}\mathrm{Cube}} = \left(
\begin{array}{ccc}
1 + \overline{x} + \overline{y} + \overline{xy} &  0 & 0\\
1 + \overline{y} + \overline{z} + \overline{yz} & 0 & 0\\
1 + \overline{x} + \overline{z} + \overline{xz} & 0 & 0\\
0 & 1 + z & 0\\
0 & 1+x & 1+x\\
0 & 0 & 1 + y
\end{array} \right)
\end{align}
This stabilizer map corresponds to the topologically ordered ``X-Cube'' fracton Hamiltonian
presented in the main text.  
Second, all commuting Hamiltonians for $\mathbb{Z}_{N}$
topological order in two- and three-dimensions 
are described by stabilizer maps of this form.
The stabilizer Hamiltonian for three-dimensional $\mathbb{Z}_{2}$ topological order is given by
\begin{align}
S_{\mathbb{Z}_{2}} = \left(
\begin{array}{cccc}
1 + \overline{x} & 0 & 0 & 0\\
1 + \overline{y} & 0 & 0 & 0\\
1 + \overline{z} & 0 & 0 & 0\\
0 & 1 + y & 1 + z & 0\\
0 & 1 + x & 0 & 1 + z\\
0 & 0 & 1 + x & 1 + y
\end{array}
\right).
\end{align}
Furthermore, the Haah's code, 
a topologically ordered state with \emph{only} immobile topological excitations 
is described by the stabilizer map
\begin{align}
S_{\mathrm{Haah}} = \left(\begin{array}{cc}
1 + \overline{xy} + \overline{yz} + \overline{xz} & 0\\
1 + \overline{x} + \overline{y} + \overline{z} & 0\\
0 & 1 + {x} + {y} + {z}\\
0 & 1 + {xy} + {yz} + {xz}
\end{array}\right).
\end{align}

\begin{figure}
\includegraphics[trim = 4 4 4 2, clip = true, width=0.43\textwidth, angle = 0.]{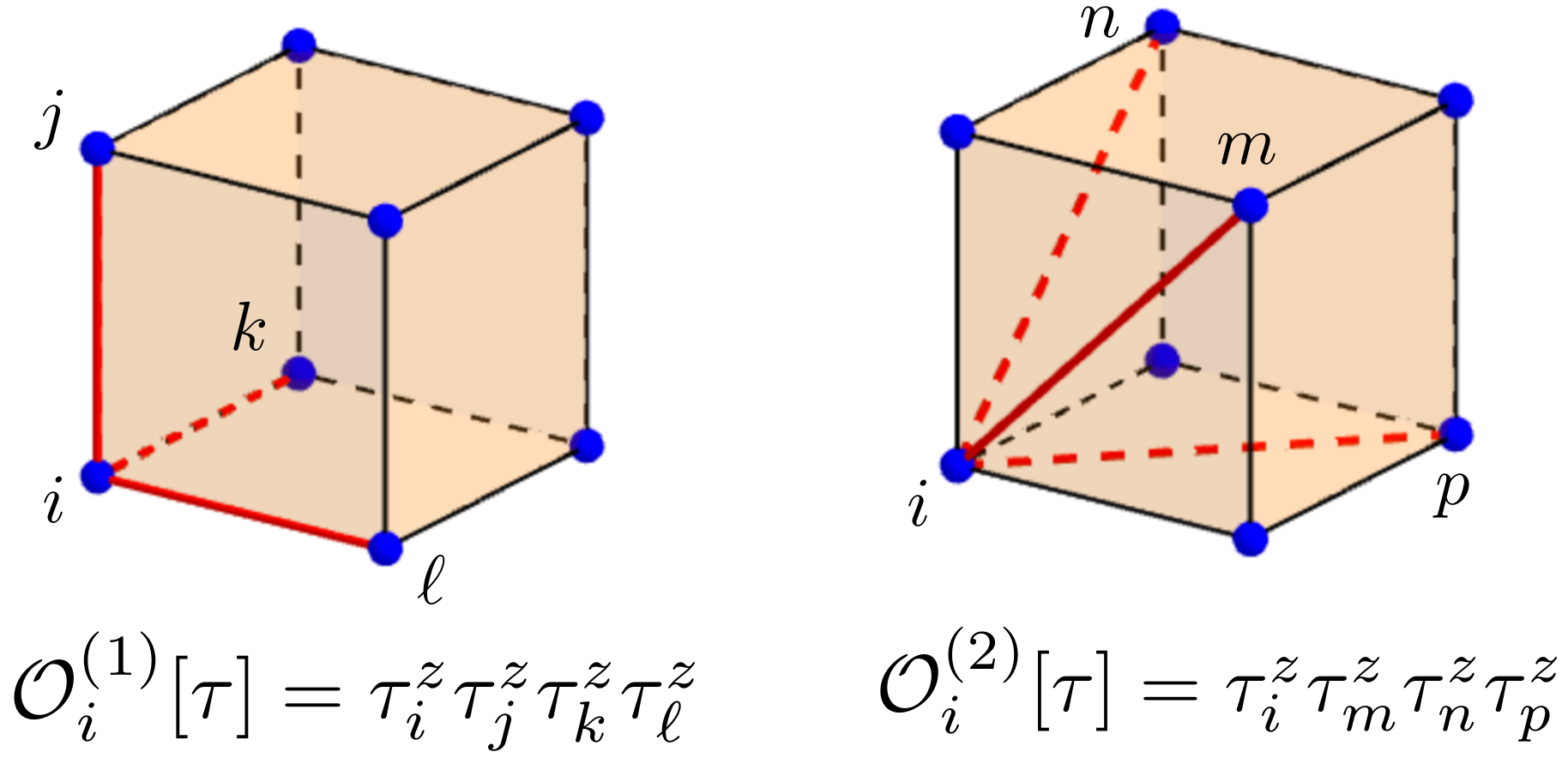}\\
\caption{
The classical spin model defined on the cubic lattice 
with fractal subsystem symmetry that corresponds to Haah's code.
The spin model may be conveniently written as 
a sum of two types of four-spin interactions at each cube, as indicated above. 
}
\label{fig:Haah_Code_Spin_Model}
\end{figure}

We may immediately read off the classical spin systems
whose symmetries have been ``gauged'' 
to obtain these topologically ordered states,
since the stabilizer map $S_{f}$ contains the \emph{classical} excitation map $E_{c} = S_c^\dagger \lambda_1$.
For example, from the stabilizer map for 3D $\mathbb{Z}_{2}$ topological order, we identify the stabilizer map
\begin{align}
S_{\mathrm{Ising}} = (1 + x, 1 + y , 1 + z)
\end{align}
which precisely describes the classical 3D Ising model on the simple cubic lattice.
A more exotic example is given by the classical spin system 
\begin{align}
S_{\mathrm{fractal}} = (\underbrace{1 + xy + yz + xz}_{f}, \underbrace{1 + x + y + z}_{g})
\end{align}
which is used to build Haah's code. 
A schematic representation of this classical spin Hamiltonian 
is shown in Fig.~\ref{fig:Haah_Code_Spin_Model} 
as a sum of two types of four-spin interactions 
at each cube on the three-dimensional cubic lattice. 
The precise subsystem symmetry in this classical spin model 
is highly sensitive to the boundary conditions imposed on the system.
For simplicity, we consider the length-$L$ three-torus with $L = 2^{n}$ 
for positive integers $n$.
Using the fact that $f^{2^{n}}$, $g^{2^{n}} \in \mathfrak{b}_{L} = ( x^L+1,y^L+1,z^L+1)$,
we find a symmetry generator
\begin{align}\label{eq:h_Haah}
h = f^{2^{n}-1}g^{2^{n}-1}.
\end{align}
This symmetry generator, 
which corresponds to a spin-flip transformation 
on a fractal configuration on the three-dimensional cubic lattice 
that resembles a Sierpinski triangle, 
is quite different from the planar symmetry for the plaquette Ising model.
It is this \emph{fractal} subsystem symmetry 
that is ``gauged" to obtain the corresponding topologically-ordered state,
known as Haah's code \cite{Haah_Code_Bib}.

\subsection{The codimension condition implies topological order}

The codimension of an ideal of polynomials $f_i$ is the codimension of the algebraic variety (zero-locus)
defined by the system of polynomial equations $f_i = 0$.
The variety may be neither smooth nor connected.
One has to look at the component of the largest geometric dimension
and take the difference of that dimension from the ambient dimension,
in order to obtain the correct codimension.
For instance, the codimension of $((x-1)(y-1),(x-1)(z-1))$ is 1
since the zero-locus contains a plane $x=1$ in three-dimensional space,
whereas the codimension of $(z-1,y-1)$ is 2 
since the zero-locus is the line given by $z=1=y$.
Although this codimension criterion can never be met in one spatial dimension,
the codimension criterion is fairly mild in higher dimensions.
For example, it is satisfied by the standard Ising model with nearest neighbor interaction
on two or higher dimensional lattices, and the ideal $I(S_c)$
has codimension that is equal to the spatial dimension,
the maximum possible value.
In fact, the exotic topological phases that we investigate
exploits this mild requirement,
and all our examples have the property that $I(S_c)$ has codimension exactly two,
saturating the inequality in the criterion $\codim I(S_c) \ge 2$.

We now return to the stabilizer map $S_{f} = E_{c} \oplus G$
and show that $S_{f}$ indeed corresponds to a topologically ordered and
commuting Hamiltonian.
First, $S_{c} \cdot G = 0$ implies that $G^{\dagger}\cdot S_{c}^\dagger = 0$,
and hence $E_{f} \cdot S_{f} = 0$;
the Hamiltonian corresponding to the stabilizer map $S_{f}$ 
consists of all commuting terms.
In addition, we have $\ker S_{c} = \im G$ by definition.
As a result, the local indistinguishability condition for any degenerate ground-states
of the stabilizer Hamiltonian on the torus,
given by $\ker E_{f} = \im S_{f}$,
is satisfied if and only if
$\ker G^{\dagger} = \im S_{c}^\dagger$.
Said differently, the complex
\begin{align}\label{eq:Complex}
0 \to R^1 \overset{S_c^\dagger}{\longrightarrow} R^{\ell} \overset{G^{\dagger}}{\longrightarrow}R^{m}
\end{align}
where $R^{n}$ is a free module of rank $n$
over the Laurent polynomial ring $R = \mathbb{F}_{2}[x^{\pm}, y^{\pm}, z^{\pm}]$,
must be exact.
The equivalent condition for this to be true is our condition $\codim I(S_c) \ge 2$.
The proof of this is purely algebraic, and a general reader might want to skip the rest of this subsection.

A criterion for the exactness of a complex of free modules 
is provided by the following \cite{Buchsbaum_Eisenbud_Bib, Eisenbud_Bib}:\\

\noindent 
{\bf Theorem} [Buchsbaum-Eisenbud]
{\it
A chain complex of free modules over $R = \mathbb F_2[x^\pm, y^\pm, z^\pm]$
\begin{align}
0 \to F_{n}\overset{\phi_{n}}{\longrightarrow}F_{n-1}\overset{\phi_{n-1}}{\longrightarrow}
\cdots F_{2}\overset{\phi_{2}}{\longrightarrow}F_{1}\overset{\phi_{1}}{\longrightarrow}F_{0}
\end{align}
is exact if and only if 
\begin{itemize}
\item $\mathrm{rank}(F_{i}) = \mathrm{rank}(\phi_{i}) + \mathrm{rank}(\phi_{i+1})$ for $i = 1,\ldots, n$ and
\item $\codim I(\phi_{i}) \ge i $ for $i = 1,\ldots, n$.
\end{itemize}
Here, $I(\phi_{i})$ is the $k$-th determinantal ideal $I_k(\phi_i)$ 
with the largest $k$ such that $I_k(\phi_i) \neq 0$.
}\\

The $k$-th determinantal ideal is one that generated by all determinants of $k \times k$ submatrices.
It is important that the complex is terminated with 0 at the left end.
(The original theorem is more general than we state here, and is given in terms of {\em depth}
of the determinantal ideal. In general, the codimension only upper bounds the depth,
and the two notions are not equal. However, our Laurent polynomial ring is sufficiently nice,
i.e., Cohen-Macaulay, that the two quantities are equal for any ideal;
our ring is obtained from a polynomial ring over a field 
followed by localization by a single element $xyz$.)

We apply the Buchsbaum-Eisenbud criterion to our complex \eqref{eq:Complex}.
Since $\ker S_c = \im G$, we know by the theorem that
$\rank S_c + \rank G = \ell$ and $\codim I(G) \ge 2$.
Since the conjugation is an automorphism of $R$, the rank is invariant under conjugate transpose.
This implies that $\rank S_c^\dagger + \rank G^\dagger = \ell$ and $\codim I(G^\dagger) \ge 2$.
Since $S_c^\dagger$ has one column,
we have $\rank S_c^\dagger = 1$.
These implies the first condition.
All we need is the second condition, $\codim I(G^\dagger) \ge 1$ and $\codim I(S_c^\dagger) \ge 2$.
The former is already shown, and the latter is exactly our requirement.
 
An example of a classical spin system that \emph{violates} our codimension condition 
is given by the stabilizer map 
$S_{c} = (f, g)$, with $f = 1 + x$ and $g = 1 + x + y + z + xy + yz + xz + xyz$.
This classical spin system has a \emph{linear} subsystem symmetry as the symmetry generator 
\begin{align}
h_x = \sum_{n \in \mathbb Z} x^n
\end{align}
satisfies $h_x f  = 0 = h_x g$.
We see that $\codim I(S_{c}) = 1$,
since the zero-locus of the polynomials $f$ and $g$ contains a plane given by $x=1$.
Therefore, the stabilizer map $S_{f} = S_{c}^\dagger \oplus G$ for this classical spin model 
corresponds to a commuting Hamiltonian 
with degeneracy due to \emph{local} observables (order parameter)
and does not exhibit topological order.

While we have provided
an algebraic condition on the classical spin model that guarantees that the dual Hamiltonian 
with stabilizer map $S_{f}$, has no local observables, we may also argue for the local indistinguishability of the ground-states using 
the \emph{F-S} duality.  
Ground-states of the Hamiltonian $H_{\mathrm{fracton}}$, using the notation presented in the main text, 
satisfy $B_{i}^{(k)}\ket{\Psi} = A_{i}\ket{\Psi} = \ket{\Psi}$, and will be locally indistinguishable if any local 
operator that commutes with $H_{\mathrm{fracton}}$ can be written as a product of nexus charges $A_{i}$ 
or generalized monopole operators $B_{i}^{(k)}$.  Recall that the generalized monopole operators are 
defined as the set of {all} local, $\sigma^{z}$-type operators that commute with the nexus charges. 
Therefore, by construction, the only $\sigma^{z}$-type operators that commute with $H_{\mathrm{fracton}}$ 
are products of $B_{i}^{(k)}$. 

We now invoke the \emph{F-S} duality to argue that any local, $\sigma^{x}$-type operator that commutes 
with $H_{\mathrm{fracton}}$ can be written as a product of nexus charges, provided that the classical spin 
model $H_{0}$ has no lower-dimensional ($d_{s} < 2$) subsystem symmetries.  Assume that such a $\sigma^{x}$-type 
operator $\mathcal{O}_{X}$ does exist, that can 
distinguish the ground-states of $H_{\mathrm{fracton}}$.  Since this operator commutes with all of the 
generalized monopole operators, $\mathcal{O}_{X}$ corresponds to a valid domain wall configuration that 
can be created by acting on the ground-state of the spin model $H_{0}$.  
Therefore, we may invoke the \emph{F-S} duality to construct 
a dual representation of $\mathcal{O}_{X}$ in terms of the Ising matter fields. Since $\mathcal{O}_{X}$ is a 
$\sigma^{x}$-type operator that cannot be written as a local product of nexus charges, however, its dual 
representation $\widetilde{\mathcal{O}_{X}}$ will be given by a \emph{non-local} product of $\tau^{x}_{i}$ 
operators. 
Physically, this duality implies that performing spin-flips along a this non-local by acting with 
$\widetilde{\mathcal{O}_{X}}$ on the classical ground-state of $H_{0}$ 
will flip a local set of interaction terms $\mathcal{O}_{i}^{(a)}$ at a point at the boundary of the region, as specified by the 
support of the operator $\mathcal{O}_{X}$.    
By taking products of the $\mathcal{O}_{X}$ operators, it is possible to reduce the dimension $d_{s}$ along which the dual operator $\widetilde{\mathcal{O}_{X}}$ has support to $d_{s} < 2$. However, this immediately implies that on the torus, the classical spin system $H_{0}$ has a subsystem symmetry along a region that is of smaller dimension than a plane.
Therefore, 
by requiring that $H_{0}$ has no lower-dimensional $d_{s} < 2$ 
subsystem symmetries, we guarantee the local indistinguishability of
the ground-states of $H_{\mathrm{fracton}}$.

\subsection{The Fracton Condition}

We now consider the excitation map $E_{f}$ for the commuting,
topologically-ordered Hamiltonian built from the data contained in the classical spin model.
We show that the topological excitations of this model are \emph{fractons}
if and only if 
the stabilizer ideal $I(S_{c}) \equiv \langle f_{1}, \ldots, f_{n}\rangle$
contains no binomial terms (the fracton condition above).
From the excitation map $E_{f}$ (\ref{eq:E_f_map}),
we observe that one of the excitations
(the nexus charge) lives on the sites of the lattice.
An operator $\mathcal{O}$ that is a product of $\sigma^{z}$ terms,
when acting on the ground-state,
generates a pattern of nexus charge excitations.
The possible nexus charge configurations by finitely supported operators
are in one-to-one correspondence with the image of the excitation map;
this is the defining property of the excitation map.
The submatrix of the excitation map that is responsible for the nexus charge
consists of one row, which is the same as $S_c$,
and therefore the image is equal to the $I(S_{c})$.

A topological excitation is a \emph{fracton}
if it is impossible to create an isolated pair of such excitations from the ground-state.
Hence, the stabilizer Hamiltonian $S_f$ exhibits fracton topological order
if and only if $I(S_c)$ contains no binomial terms
\begin{align}
1 + x^{n_{1}}\cdots x^{n_{d}} \notin I(S_{c})
\end{align}
for all $n_{1}$, $\ldots$, $n_{d} \in \mathbb Z$.
This is a condition on the classical spin Hamiltonian
that no product of any interactions $h^{(i)}[\tau]$ can simply be a product of a \emph{pair} $\tau^{z}$ terms.
For example, the classical Ising model, with nearest-neighbor spin-spin interactions,
trivially violates this condition in any dimension.
It is not always obvious to test whether an ideal contains a binomial.
We explain one test technique with examples shortly.

The fracton condition cannot be satisfied
if the codimension of $I(S_c)$ is equal to the spatial dimension,
the maximum value~\cite{Polynomial_Bib}.
In the latter is true, then any point-like excitation
appears at the tip of some string like operator, which can be freely
bent in the system, and it behaves as an ordinary particle.

\subsection{Classical Spin Models with Planar subsystem Symmetry}

In this section, 
we show that applying our construction to classical spin models 
with spin-flip symmetries along planes always yields a fracton topological phase.
Consider a stabilizer map 
$S_c = (f_{1},\ldots, f_{n})$
for a classical spin Hamiltonian,
where $f_1, \ldots, f_n$ are Laurent polynomials.
As we have shown in Appendix~\ref{sec:symmetry},
the symmetry generators
\begin{align}
h_{xy} &= \sum_{n,m \in \mathbb Z} x^n y^m ,\nonumber \\
h_{yz} &= \sum_{n,m \in \mathbb Z} y^n z^m ,\label{eq:PlanarSymmetry}\\
h_{xz} &= \sum_{n,m \in \mathbb Z} x^n z^m  \nonumber
\end{align}
satisfy
\begin{align*}
h_{xy} f_{i} &= 0\\
h_{yz} f_{i} &= 0\\
h_{xz} f_{i} &= 0
\end{align*}
for all $i = 1,\ldots, n$.
We also have seen that $h_{xy} g(x,y,z) = g(1,1,z)$, etc., for any Laurent polynomial $g$.
The symmetry condition then implies that the polynomials $f_i(1,1,t), f_i(1,t,1), f_i(t,1,1)$ are
identically zero for all $i$. That is, $(1,1,t)$, etc., are roots of the polynomials $f_i$.
Geometrically, this means that the variety defined by $f_i(x,y,z) = 0$
contains three lines parametrized by $t \mapsto (1,1,t), (1,t,1), (t,1,1)$.
 
We can use this observation to show the fracton condition.
Suppose on the contrary that the ideal $I(S_c)$ contains a binomial term.
\begin{align}
 1 + x^a y^b z^c \in I(S_c).
\end{align}
Since any member of the ideal is a linear combination of generators,
we have
\begin{align}
 1+ x^a y^b z^c = \sum_i g_i(x,y,z) f_i(x,y,z) .
\end{align}
Let us substitute $x,y,z$ with the solution $(1,1,t)$:
\begin{align}
 1 + t^c = \sum_i g_i(1,1,t) f_i(1,1,t) = 0.
\end{align}
Since this has to be true as a polynomial in $t$, we deduce that $c = 0$.
Likewise, the solution $(1,t,1)$ implies $b=0$, and $(t,1,1)$ implies $a=0$.
Therefore, our binomial was actually $1 + x^0 y^0 z^0 = 0$,
and the ideal $I(S_c)$ does not contain any binomial.

It has to be noted that the fracton condition (the absence of binomial) alone
does not imply that the F-S dual of the classical spin model is topologically ordered;
the codimension condition should be checked separately.
If the initial classical model had line-like symmetries,
then we would have solutions of form e.g. $(t_1, t_2, 1)$.
This has a codimension 1, and the hence $\codim I(S_c) \le 1$,
and the F-S dual would not be topologically ordered.

We conclude that for any classical spin system 
with spin-flip symmetries along all three orthogonal planes and no ``lower dimensional'' symmetries,
our construction for gauging this symmetry 
will give rise to a fracton topological phase.

Let us demonstrate that the nexus charge in the ``X-Cube" Hamiltonian
is a fracton excitation using the above technique.
Recall that 
$I(S_{c}) = \langle 1 + x + y + xy, 1 + y + z + yz, 1 + x + z + xz\rangle$
for the plaquette Ising model.
The symmetry generators $h_{xy}, h_{yz}, h_{xz}$ of \eqref{eq:PlanarSymmetry}
is indeed the symmetries of the X-Cube model, and therefore $I(S_c)$ does not contain any binomial.
By solving the equation $1 + x + y + xy=0$, $1 + y + z + yz=0$, and $1 + x + z + xz=0$,
we see that $(1,1,t)$, $(1,t,1)$, and $(t,1,1)$ are the only solutions,
which are lines of geometric dimension 1,
and therefore the codimension of $I(S_c)$ is equal to 2.
Therefore, X-Cube model has fracton (immobile) topological excitations.
This derivation provides a formal proof of the simple physical statement
that no product of the four-spin interactions in the classical plaquette Ising model
can act exclusively on an isolated pair of spins.

\begin{figure}
$\begin{array}{cc}
 \includegraphics[trim = 4 4 4 2, clip = true, width=0.14\textwidth, angle = 0.]{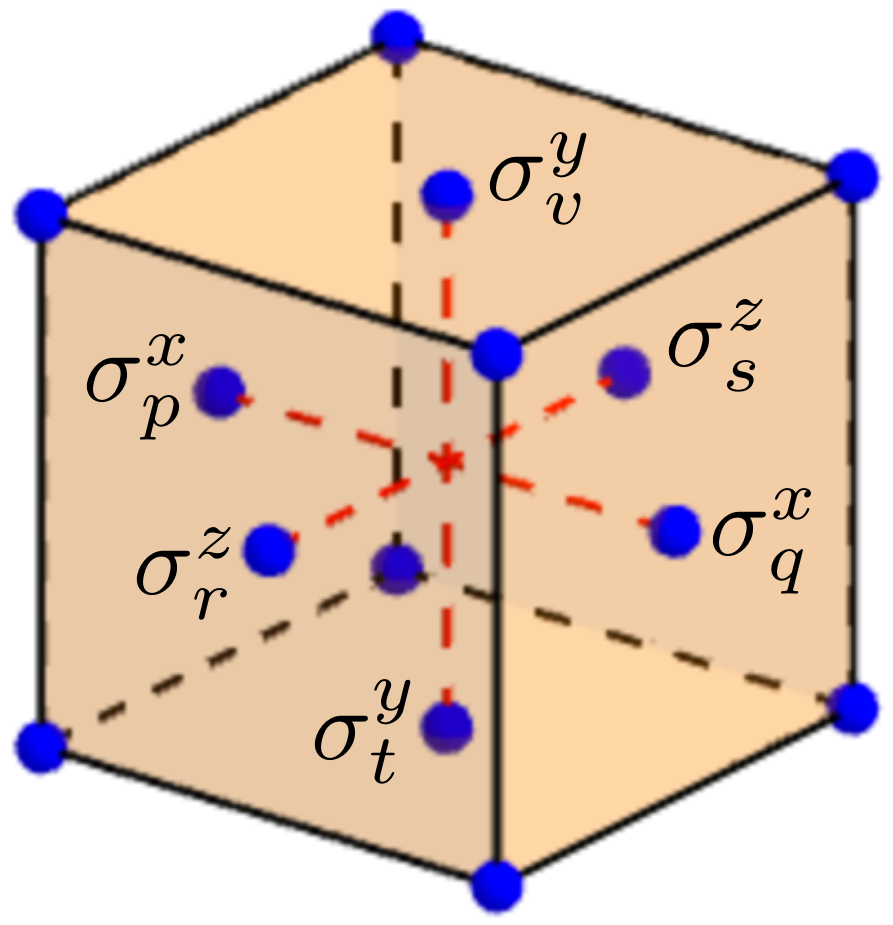}\,\,\,
 & \,\,\,\, \includegraphics[trim = 4 4 -40 2, clip = true, width=0.29\textwidth, angle = 0.]{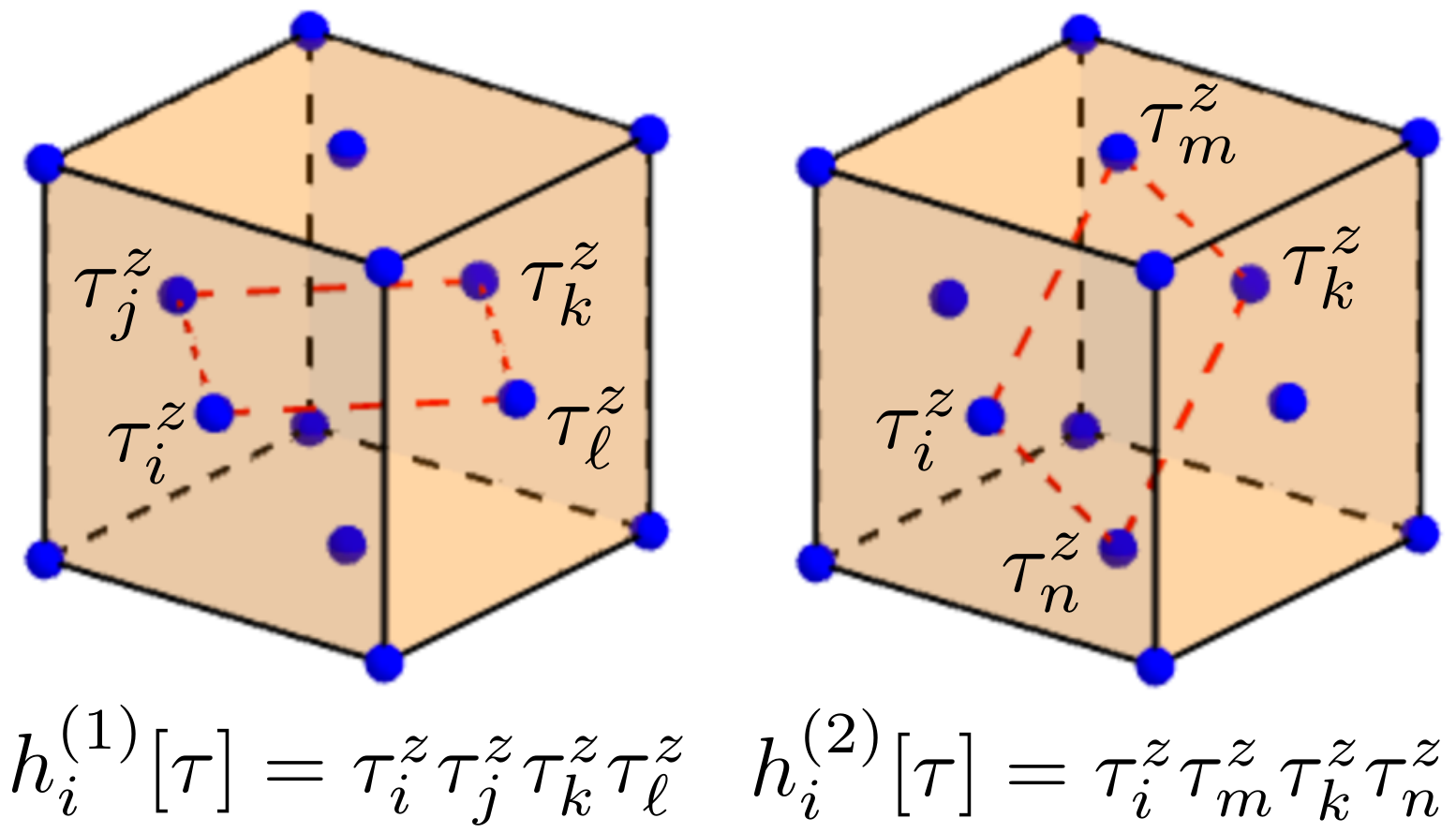}\\
 \text{CBLT Model} & \\
 &\\
 \text{(a)} & \text{(b)}
 \end{array}$
\caption{
The (a) CBLT model represented on an fcc lattice.
The model consists of a single, six-spin interaction term per lattice site.
This model may be constructed from the interacting spin model shown in (b) on the fcc lattice,
with two four-spin interactions per lattice site.
}
\label{fig:Chamon_Model}
\end{figure}

We can adapt the above argument to construct
another spin model with fracton topological order.
Consider the classical spin system
\begin{align}
S_c = (1 + x + z + x\overline{z}, \,\,1 + x + y + x\overline{y})
\end{align}
Observe that $f_{1} = 1 + x + z + x\overline{z} = \overline{z}(1 + z)(x + z)$ 
and $f_{2} = 1 + x + y + x\overline{y} = \overline{y}(1 + y)(x+y)$.
Thus we identify the three planar symmetries of this classical spin model
\begin{align}
h_{1} &= \sum_{n,m \in \mathbb Z} x^{n}y^{m},\\
h_{2} &= \sum_{n,m \in \mathbb Z}(x+z)^{n} (x+y)^{m},\\
h_{3} &= \sum_{n,m \in \mathbb Z} x^{n} (x+z)^{m}.
\end{align}
It is routine to check that $h_i f_j = 0$.
Since $f_{1}(x,y,z)$ and $f_{2}(x,y,z)$ do not have any common factor,
the ideal $I(S_c) = \langle f_1, f_2 \rangle$ has codimension 2,
satisfying our codimension condition.
The set of relations between $f_1$ and $f_2$ (the kernel of $S_c$),
is generated by 
$G = (1 + \overline{x} + \overline{y} + \overline{x}{y},\,1 + \overline{x} + \overline{z} + \overline{x}{z})^{T}$. 
Through our F-S duality, we obtain a stabilizer map
\begin{align}
S_{\mathrm{CBLT}} = \left(
\begin{array}{cc}
1 + x + z + x\overline{z} & 0\\
1 + x + y + x\overline{y} & 0\\
0 & 1 + \overline{x} + \overline{y} + \overline{x}{y}\\
0 & 1 + \overline{x} + \overline{z} + \overline{x}{z}
\end{array}
\right),
\end{align} 
which corresponds to a stabilizer Hamiltonian with fracton topological order.
In fact, this stabilizer map has an extra property that $f_1 \bar f_2 + \bar f_1 f_2 = 0$,
which allow us to consider a simpler stabilizer map
\begin{align}
S_{\mathrm{CBLT}} = S_c^{T} = \left(
\begin{array}{c}
1 + x + z + x\overline{z}\\
1 + x + y + x\overline{y}
\end{array}
\right)
\end{align} 
corresponding to a fracton phase with ``half'' of the topological degeneracy.
This stabilizer map is precisely the CBLT model \cite{Chamon, Bravyi},
which has a single type of fracton excitation,
and degeneracy $D$ given by $\mathrm{log}_{2}D = 8L$ \cite{Bravyi,Polynomial_Bib}
on the length-$L$ three-torus.
The CBLT model is conveniently represented on the face-centered cubic
(fcc) lattice as shown in Figure \ref{fig:Chamon_Model}a.

\subsection{Planar Symmetry and Classical Spin Models with an $m$-Site Unit Cell}

So far we have assumed that a classical spin model
had a single spin per Bravais lattice site.
When the unit cell has $m$ spins ($m \ge 1$),
our duality construction and criteria for topological order
directly carry over.
However, the fracton condition and its relation to planar symmetries
require some extra attention.

The stabilizer map $S_c$ for the classical spin ($\tau^z$) system
is an $m \times \ell$ matrix of Laurent polynomials 
(omitting the lower half-block representing $\tau^x$),
where $\ell$ is the number of interactions per unit cell.
\begin{align}
S_{c} = \left(
\begin{array}{cccc}
f_{1}^{(1)} & f_{2}^{(1)} & \cdots & f_{\ell}^{(1)}\\
f_{1}^{(2)} & f_{2}^{(2)} & \cdots & f_{\ell}^{(2)}\\
\vdots & \vdots & \ddots & \\
f_{1}^{(m)} & f_{2}^{(m)} & \cdots & f_{\ell}^{(m)}
\end{array}\right)
\end{align}
Our generalized gauge theory prescription produces
a (quantum) stabilizer Hamiltonian described by
$ S_{f} \equiv S_{c}^{\dagger} \oplus G $
where 
$S_{c}^{\dagger} = (\overline{S_{c}})^{T}$ and 
$G$ is a matrix of the generators of $\ker S_c$.
The matrix $G$ is nonzero if and only if $k = \rank S_c < \ell$.
This condition is satisfied when $S_{c}$ 
is a row matrix and there are two or more interactions per spin,
as is the case in all the examples in this paper
with a single site per unit cell.
The Buchbaum-Eisenbud criterion for exact sequence can be applied,
and, if $G \neq 0$,
we see that the only nontrivial condition is that $\codim I(S_{c}) \ge 2$
where $I(S_c) = I_k(S_c)$ is the $k$-th determinantal ideal with the largest $k$ such that $I_k(S_c)$ is nonzero.
Summarizing, the topological order is realized in $S_f$ if and only if
\begin{itemize}
 \item $k = \rank S_c < \ell$, and 
 \item $\codim I_k(S_c) \ge 2$.
\end{itemize}

We now assume that the classical system has planar subsystem symmetry.
The three independent generators of spin-flip transformations along orthogonal planes
are denoted $h_{xy}$, $h_{yz}$ and $h_{xz}$.
The classical system on a lattice with an $m$-site basis has planar subsystem symmetry if and only if
\begin{align}
h_{j} \cdot \sum_{i=1}^{m} f_{k}^{(i)}(x,y,z) = 0
\end{align}
for all $j = xy,yz,zx$ and $k = 1,\ldots,\ell$.
As before, this implies that
\begin{align*}
  \sum_{i=1}^{m} f_{k}^{(i)}(1,1,t) &= 0,\\
  \sum_{i=1}^{m} f_{k}^{(i)}(1,t,1) &= 0,\\
  \sum_{i=1}^{m} f_{k}^{(i)}(t,1,1) &= 0.
\end{align*}

Suppose the nexus charge exists for the moment.
The condition that the nexus charges, the excitation of $S_c^\dagger$ terms in $S_f$,
are all fractons is that $\im S_{c}$ does not contain any two-term element of form
\begin{align*}
 \begin{pmatrix}
 \vdots \\ 0 \\ 1 + x^a y^b z^c \\ 0 \\ \vdots
 \end{pmatrix}.
\end{align*}
If there is such a two-term element,
then
\begin{align*}
 \begin{pmatrix}
 \vdots \\ 0 \\ 1 + x^a y^b z^c \\ 0 \\ \vdots
 \end{pmatrix} = 
 \begin{pmatrix}
  \vdots \\ 
  \displaystyle\sum_{k=1}^\ell f_k^{(i-1)} (x,y,z) g_k(x,y,z) \\
  \displaystyle\sum_{k=1}^\ell f_k^{(i)} (x,y,z) g_k(x,y,z) \\
  \displaystyle\sum_{k=1}^\ell f_k^{(i+1)} (x,y,z) g_k(x,y,z) \\
  \vdots
  \end{pmatrix}.
\end{align*}
Summing all the components, we get
\begin{align}
 1 + x^a y^b z^c = \sum_{k,i} f_k^{(i)} g_k(x,y,z).
\end{align}
Evaluating the polynomial at $(1,1,t)$, etc., we conclude that $a=b=c=0$.
That is, there is no two-term.

The existence of the nexus charge might seem mundane,
but we do not know whether this is true even assuming the various conditions
we have discussed.
It is however known that in three spatial dimensions
any topologically ordered stabilizer Hamiltonian
has a point-like excitation~\cite{Polynomial_Bib}.
This implies that either the nexus charge sector or the nexus flux sector
admits isolated point-like excitations.
If this point-like excitation happens to be a nexus charge,
then the symmetry of the classical side implies that this charge is immobile (fracton).
We note that when the unit cell consists of a single spin in the classical side ($m=1$),
the existence of the nexus charge was immediate since $\coker S_c$ is a torsion module.
We conclude this section by summarizing our discussion.
\begin{itemize}
 \item Given an aribitrary classical spin model with planar symmetries,
 with the number of interaction terms per Bravais lattice site is greater than
 the size of the unit cell,
 any nexus charge, if exists, is a fracton.
\end{itemize}

\section{Ground-State Degeneracy of the Checkerboard and X-Cube Models } 

Here,
we derive the ground-state degeneracy of the checkerboard and X-cube model,
constructed from the tetrahedral Ising and plaquette Ising models, respectively.

Let us briefly review the properties of the checkerboard model.
Consider a three-dimensional cubic lattice of spin-1/2 degrees of freedom.
Each spin sits at the vertex of a cube,
with the cubes arranged in a checkerboard configuration,
so that any pair of neighboring cubes overlap 
on exactly two spins as shown in Table~\ref{fig:Classical_Spins}.
We define the Hamiltonian
\begin{align}\label{eq:H_Fracton}
H_{0} = -\sum_{c}\prod_{n\in\partial c}\sigma^{x}_{n} - \sum_{c}\prod_{n\in\partial c}\sigma^{z}_{n}.
\end{align}
where each product is taken over the eight spins sitting at the vertex of cube $c$,
while the sum is taken over cubes in the checkerboard configuration.
The topological excitations in this model
are similar to that of the Majorana cubic model~\cite{Fracton_Bib}.
The fundamental fracton excitation may be isolated at the corners of a membrane operator,
given by the product of spin operators along a flat, two-dimensional surface.
Composites of fracton excitations, however,
are topological excitations that are free to move
along one- and two-dimensional subsystems without any energy cost.
Wilson line operators, given by the product of spin operators along \emph{straight} lines,
create pairs of fracton excitations at each end;
these composite excitations are  ``dimension-1'' particles,
which are only free to move along the Wilson line.
A pair of parallel Wilson lines in adjacent layers, however,
maybe used to create excitations that are free to move within the plane.
These topological excitations are \emph{anyons} with well-defined mutual and self-statistics.   

The stabilizer map for the checkerboard model is given by the expression
\begin{align}
S_{\mathrm{check}} = 
\begin{pmatrix}
f & 0\\
&\\
\overline{f} & 0\\
&\\
0 & \overline{f}\\
&\\
0 & f 
\end{pmatrix}
\end{align}
where $f = 1 + x + y + z \in \mathbb{F}_{2}[x^{\pm}, y^{\pm}, z^{\pm}]$.

For a topologically ordered stabilizer Hamiltonian
with the stabilizer map $S$ and excitation map $E = S^\dagger \lambda_\ell$,
the ground state degeneracy $D$ on a length-$L$ torus
is given by 
\begin{align}
 &2\log_2 D = \dim_{\FF_2} \ker E / \im S \quad \nonumber\\
 &\text{over}\,\,\,\FF_2[x,y,z]/\langle x^L+1, y^L+1, z^L+1 \rangle .
\end{align}
If the number of spins in the unit cell, $\ell$, is equal to the number of interaction types
(the number of columns in $S$), then 
\begin{align}
 \log_2 D = \dim_{\FF_2} \coker S^\dagger .
\end{align}
This is applicable for all examples in this paper,
except for the $\mathbb Z_2$ gauge theory in three-dimensions.
In our examples where $S^\dagger = S_c \oplus G^\dagger$ is block-diagonal,
the formula further decomposes as
\begin{align}
 \log_2 D = \dim_{\FF_2} \coker S_c + \dim_{\FF_2} \coker G^\dagger.
\end{align}

Since the checkerboard stabilizer map has an extra structure that $S_c = G^\dagger$,
the degeneracy calculation reduces to that of 
\begin{align*}
 \log_2 D_\text{check} = 2 \dim_{\FF_2} \frac{\FF_2[x,y,z]/\langle x^L+1,~ y^L+1,~ z^L+1 \rangle}{\langle 1 + x + y + z,~xyz+xy+yz+zx \rangle} .
\end{align*}
This was calculated in Ref.~\cite{Fracton}, and the answer is $\log_2 D_\text{check} = 6L -6$.

The X-Cube model has different blocks in $S_\text{X-Cube}^\dagger$.
Here we calculate the degeneracy of $H_\text{X-Cube}$ for odd $L$ only.
One component reads
\begin{align}
 \frac{\FF_2[x,y,z]/\langle x^L+1,~ y^L+1,~ z^L+1 \rangle}{\langle (1+x)(1+y),~(1+y)(1+z),~(1+x)(1+z) \rangle}.
\end{align}
One may extend the coefficient field to the algebraic closure $\FF$,
and since $L$ is odd, $x^L+1 =0$ has $L$ distinct roots, one of which is $x=1$.
Localizing at the maximal ideal $\langle x+t, y+1, z+1 \rangle$,
we see that the factor ring becomes just $\FF$ of dimension 1.
Using the cyclic symmetry $x \to y \to z \to x$, 
we see that the factor ring has dimension $3(L-1) + 1 = 3L -2$.
The other component is
\begin{align}
 \frac{(\FF[x,y,z]/\langle x^L+1,~ y^L+1,~ z^L+1 \rangle)^2}{\begin{pmatrix}
z+1 & 1+x & 0 \\
0 & 1+x & 1+y
 \end{pmatrix}}
\end{align}
The second determinantal ideal of the matrix in the denominator is the same as $I(S_c)$,
so we only have to consider localization at points $(1,1,t)$, $(1,t,1)$, and $(t,1,1)$.
Localization at $(1,1,t \neq 1)$ amounts to evaluating the matrix at that point due to the boundary conditions,
and the factor module becomes $\FF^1$.
At $(1,1,1)$, the matrix becomes zero, and the factor module becomes $\FF^2$.
Hence, the factor ring has dimension $3(L-1)+2 = 3L-1$.
The degeneracy $D_\text{X-Cube}$ is thus given by
\begin{align}
 \log_2 D_\text{X-Cube} = 6L -3 \text{ for odd } L.
\end{align}


\begin{thebibliography}{1}
\bibitem{Footnote} More precisely, a configuration of Ising spins may be specified exactly by the eigenvalues of interaction terms in the Hamiltonian, only up to spin-flips performed along subsystems that leave all of the interaction terms invariant. 
\end{thebibliography}
\end{document}